# Generative AI and the transformation of Work in Latin America - Brazil

Carmen Bonifácio[†], Fernando Schapachnik[‡] , Fabio Porto [†]

Laboratório Nacional de Computação Científica, Instituto de Inteligência Artificial, LNCC, Brazil[†]
Universidad de Buenos AIRES, Fundación Sadosky, Argentina [‡]

## Executive Summary

This survey is a contribution for the *GPAI Future of Work* working group. It is part of the project: Voices of Change: Generative AI and the Transformation of Work in Latin America, and explores the impact perceived by employers and employee of GenAI in their work activities in Brazil.

Generative AI (GenAI) is gradually transforming Brazil's workforce, particularly in micro and small businesses, though its adoption remains uneven. This survey examines the perceptions of employers and employees across five sectors: Sales, Customer Service, Graphic Design or Photography, Journalism or Content Production, and Software Development or Coding. The results are analyzed in light of six key dimensions of workforce impact. The findings reveal a mix of optimism, apprehension, and untapped potential in the integration of AI tools.

Employees report that GenAI has helped **reduce time spent on tasks**, though this benefit is less pronounced for those working fixed hours. While professionals describe their working conditions as favorable for AI adoption, actual implementation remains limited. Concerns about job displacement persist, with **11% of respondents** citing workforce reductions and many expressing anxiety over AI's long-term effects on employment.

A significant gap exists in **training and organizational support**. Most companies have yet to prioritize upskilling, leaving employees to learn GenAI tools independently—often outside work hours. Employers attribute this delay to uncertainties around benefits, data security, and implementation strategies. However, a small but growing number of firms are considering AI adoption, signaling potential growth in Brazil's SME sector.

To ensure a smooth transition, businesses should **provide structured training, establish clear AI policies, and address employee concerns transparently**. Policymakers and industry leaders must collaborate to promote **ethical AI integration**, balancing innovation with workforce protection. Future research should broaden participation to enhance statistical robustness while fostering partnerships that incentivize engagement.

As work undergoes profound changes, aligning **human skills with technological advancements** will be crucial. This study serves as a foundation for developing **inclusive strategies** that maximize AI's benefits while safeguarding workers' rights. The IIA-LNCC supports open research and remains committed to shaping a future where technology and human potential progress together.

# Table of Contents

# 1. Introduction

Generative Artificial Intelligence (GenAI) is a type of artificial intelligence that, through training with large volumes of data, is able to create new content, such as text, images, paintings or even music. GenAI has emerged as a disruptive technology that profoundly transforms the shape of society. New systems created on its basis, change concepts and generate an impact on a range of activities whose practices have been perpetuated throughout the history of humanity. One of the sectors that has felt these interferences strongly is the labor sector. As it is the basis of society's sustainability, it requires dedicated attention to understand the direction that is being followed and the discoveries already assimilated in order to adapt.

Tasks that were primarily performed by humans are now performed, in a similar way, by a machine. A very representative example, which is evolving very quickly, is language models. In their first versions, presented to the public at the end of 2022, they were able to understand human language and integrate it with a certain cohesion only through writing. They had limitations in their construction that were quickly overcome and today, these models understand messages through voice, even differentiate between more than one interlocutor in the same conversation and also respond by voice, with greater coherence. They also store the subjects of previous conversations and can be queried even after disconnection for a long time. They are able to translate into several languages, create textual elements, pictures and sounds, speed up searches for any content available on the web, among a host of other skills that compete with human performance. This brings benefits and challenges at the same time. As benefits, we can cite the gain in personal time, the fluidity of ideas and the increase in creativity; as challenges, there are problems related to the generation of fake material, the perpetuation of social biases captured through training data, lack of responsibility, security, ethics, regulations and, above all, the impact on the labour market. For the potential of GenAI to reflect favorably on the whole of society, without deepening existing inequalities, the workforce must understand its role in this dynamic and be prepared to lead this transformation.

The GenAI is already changing the labour process and its potential can be viewed in two ways. The positive way highlights the assistance in routine activities, while the negative way expresses concern about the possibility of replacing human labor. Although some jobs will be lost, new ones will emerge. To avoid or minimize damage to the workforce, well-targeted political action will need to be taken. Some guidelines, such as encouraging learning and establishing greater contact with technology, may enable professionals to deal with the new demands.

The free availability of some GenAI applications does not imply universal accessibility, especially when it comes to users in developing countries where there is great inequality in access to technological innovations, in addition to differences in the scope of education and instruction. Another relevant issue concerns the expertise in creating these models; they are trained with data that reflect different cultures and therefore may provide responses that are not consistent with the local reality. In addition, there are risks associated with the adoption of practices that are still in constant development, aggravated by the lack of regulation on the responsibilities for the generated results. For a broad assessment of the impacts, the different segments of society must be taken into account, so that each location will present its own reality in relation to the innovations. In this sense, studies that shed light on this diversification of conditions are of utmost importance. The results contribute to the formulation of strategies increasingly focused on minimizing negative or undesirable effects. Case studies of leading companies that seek to add value to their production or service process, such as *Artificial IntelligenceI and the Labour Market* by OECD [1], *Transformation of the Warehouse Sector through AI* at Amazon [2] and *The Workers Behind AI* at Sama [3] seek to identify bottlenecks or express points of success to support and guide decision-makers.

Today, transformations are so rapid that they defy human capacity for assimilation. Faced with this rapid rise, GenAI has pushed society to disseminate the concepts and premises of ethical, responsible and safe AI, which involves transparency, equity, safety and respect for human rights when building systems. As these guidelines are followed with greater criteria of fairness and responsibility, trust in the results delivered by this technological aspect tends to increase. This increases the engagement of all those involved in the production chain, primarily promoting human well-being.

The *Global Partnership on Artificial Intelligence* (GPAI) initiative, now part of the *Organization for Economic Cooperation and Development* (OECD), establishes the "Future of Work" as one of the pillars of its agenda for the implementation of human-centered, safe and trustworthy AI. The Artificial Intelligence Institute of the National Laboratory for Scientific Computing[1] (IIA-LNCC), in Brazil, funded by the Ministry of Science, Technology and Innovation (MCTI), and the Fundación Sadosky, in Argentina, cooperated in developing a sample survey entitled "The impact of the use of Generative Artificial Intelligence in the labor market in Brazil". As a continent-sized country, Brazil has different characteristics in each of its macro-regions. This condition reveals the importance of knowing the impact of generative AI on its labor market, as it can serve as a benchmark for several other countries with similar characteristics. We were pioneers in implementing a sample survey of this size, with the aim of understanding the specific nuances of generative AI on the Brazilian labor market. We hope to make a contribution by identifying relevant points that can be improved, seeking improvements for Brazil and Latin America. The findings of this first survey will be presented throughout this report.

The remaining of this report is organized as follows. Section 2 we discuss global initiatives concerning the impact of GenAI in the workforce. Next, Section 3 presents the context of the FOW initiative in Brazil. Then, Section 4 presents the methodology, the steps taken from the design of the research to the preparation of the final report. Then, Section 5 presents the data analysis and the initial observations. Section 6 presents actions for the future and Section 7 Conclusion.

## 2. Future of Work Initiatives

Several global organizations have been involved in understanding and anticipating the impact of the advent of IAGen on the global labour market. The relationship between the workforce and the new technology is being investigated from various angles in order to devise human-centered strategies that minimize its negative impacts. This section highlights actions that address the impact of GenAI in the labor market.

### 2.1. Global Partnership on Artificial Intelligence

In the context of the GPAI initiative, the *Future of Work* working group has been set up to investigate the impact of the IAGen in the workforce. In its reports questions have been raised about the need to carry out a critical analysis of how AI can affect workers and their working environment, considering aspects such as:

- **Preparing for the Future of Work:** Discussing strategies for employers and workers to adapt to AI-driven changes.

- **Preserving and improving job quality:** Addressing concerns about job quality, inclusivity, and worker health and safety in the age of AI.

---

[1] https://instituto.ia.lncc.br/pt

- **Education and training:** Highlighting the importance of equipping the workforce with the necessary skills to thrive in an AI-powered future.
- **AI as an empowering tool:** Exploring how AI can be used to enhance workers' capabilities and productivity.

A practical study on the implementation of AI and its impact can be highlighted. The study uses as a framework the Fairwork AI principles developed in collaboration with the GPAI. The first focused at Amazon Company (*Fairwork Amazon Report 2024: Transformation of the Warehouse Sector through AI*) [1] and second at Sama Company (*Fairwork AI Ratings 2023 - The Workers Behind AI at Sama*) [2].

In the framework of the *GPAI Future of Work* working group activities, a project proposal led by Fernando Schapachnik from Fundación Dr. Manuel Sadosky, in Buenos Aires, investigated the "*Impact of generative AI on the labor market in South America*". The survey examined the impact of generative AI in Colombia, Chile, Argentina, Mexico, and Costa Rica. Our survey is a component of this study focusing on perception of the impact of GenAin in the Brazilian workforce.

## 2.2. Organization for Economic Cooperation and Development

The OECD.AI is an observatory to ensure policies, data and analysis for trustworthy artificial intelligence. The *Future of Work* is a priority for the OECD, which has joined forces with the GPAI in pursuit of human-centered, safe and reliable AI. In addition to this topic, they contribute specifically to GenAI, Data Governance, Data Privacy and AI Risk and Accountability. The organization presents a catalog with tools and metrics to ensure reliable AI ( https://oecd.ai/en/catalogue/overview ) [4].

In the context of the FOW, OECD has evaluated the impact of AI in the labor Market in a 2023 report, on the sectors of industrial manufacturing and finance ( The impact of AI on the workplace: Main findings from the OECD AI surveys of employers and workers) [5], as well as a second report (AI And The Labor Market) [1]. The OECD AI Principles for responsible stewardship of trustworthy AI, adopted in May 2019 by member countries, forms the basis for the G20 AI Principles.

## 2.3. United Nations

The United Nations (UN) has not produced specific studies about IAGen, as of the time of this report. However in October 2023 it launched an advisory body on Artificial Intelligence that brings together experts with deep experience in the government, private, technology, civil society and academic sectors, to support it in its efforts to ensure that AI is used for the greatest good of humanity. The Advisory Body is composed of 39 experts from across the world. Membership is gender-balanced, geographically diverse, and multigenerational. The multi-stakeholder High-level Advisory Body on AI does not discuss the *Future of Work* directly, but carries out analyzes and makes recommendations for the international governance of artificial intelligence as shown in Governing AI for Humanity Final Report [6].

## 2.4. European Union

In June 2021, the European Union found evidence on the expected impact of Artificial Intelligence (AI) on employment. It discusses the potential of AI in creating decent Jobs, and explores the extent to which AI offers opportunities and poses risks to working conditions. The EU board analyzed current EU and member state policies regarding the impact of AI in the working conditions. (Improving working conditions using Artificial Intelligence). The study bases its results in a targeted literature review (including 25 sources) and five semi-structured interviews with key informants to map out the current landscape of research [7].

This is a not necessarily complete view on the globally initiatives focused on the impact of AI on the labor market. Our study adds yet another strand of contribution in evaluating the impact in the workforce of GenAI in a country as geographically vast as Brazil and regionally unequal.

## 2.5. Future of Work in Brazil

According to data presented in June 2024 by the Ministry of Labor and Employment (MTE) and the Inter-Union Department of Statistics and Socioeconomic Studies (Dieese), the unemployment rate in Brazil is 8% [8]. The Brazilian Institute of Geography and Statistics (IBGE), through survey data from the quarterly series of the Continuous National Household Sample Survey (PNAD Contínua) for the same period, indicates an unemployment rate of 6.9% [9]. For the PNAD, the workforce considered unemployed is characterized in red according to the criteria presented in Figure 1.

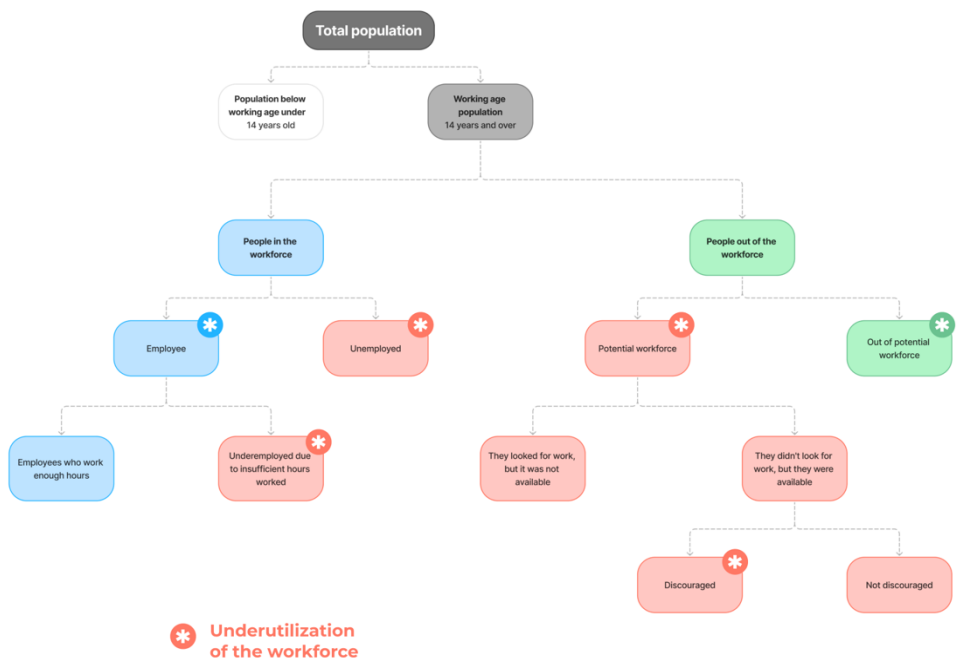

Figure 1. Workforce characterization by IBGE adapted from [10].

The surveys cover both formal and informal work (informal employment or informal economy) and can be segmented, for example, by region, gender, age group and level of education, allowing a more complete assessment of the relationships in this local and temporal context. Due to the emergence of GenAI at the end of 2022, the fluctuations and trends observed in the characterization of unemployment have received greater attention.

With the main objective of assessing workers' and employers' perceptions of the impact of GenAI on the labour market in Brazil, we decided to run an online survey. The latter targeted the impact of GenAI on a list of professional occupations, the same list used in the initiative developed in the GPAI-FOW project led by Argentina.

Our methodology considered running a sample online survey aiming at reaching a relevant part of the country. Key stakeholders associated with employers or employees unions involved with activities associated with the professional occupations of interest were contacted and invited to contribute as focal points to the survey distributing and advertising the survey among their union members.
Through the survey, we aim to capture the participants' difficulties, limitations, benefits, losses and successes as they relate to everyday working practices.
This research also aims to provide participants with moments of reflection on the applicability of GenAI tools to support the performance of their daily work tasks, raising questions about skills formation, inequalities in access and knowledge, appreciation of human beings, among other relevant aspects.

The selected professional occupations have been identified in several publications as those more susceptible toGenAI impact, measured in terms of the risk of GenAI performing tasks currently performed by professionals. The professional occupations considered in this study were:

| Software Development/Coding |
| --- |
| Graphic Designer |
| Customer Service |
| Sales |
| Journalism/Content Production |

Table 1. Professional occupations of interest.

The surveys aim to highlight the current scenario for these professions in the five regions of the country, as in Figure 2: North, Northeast, Central-West, South, and Southeast. The regions showcase the diversity of the country in terms of their environment, geography, population density, economic activity, household income, and so on.

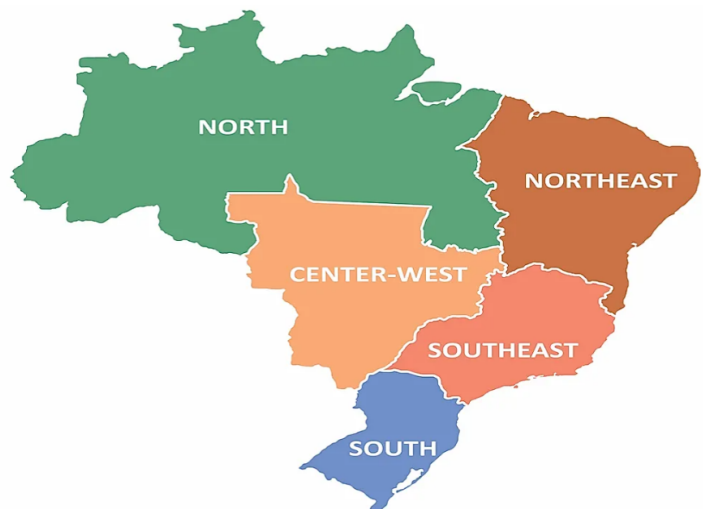

Figure 2. Macro Regions of Brazil [11].

To understand the current behavior of the sample population, the evaluation covered the following dimensions:

- Working hours
- Working conditions
- Perception of changes/inequalities
- Education and skills
- Wage
- Future

The questions composing the survey were inspired by the ones appearing in similar surveys, such as the *Stack Overflow* survey of software developers [12], the biennial technology and innovation survey *PINTEC Semestral* [13], conducted by IBGE, and a study on *The impact of AI on the workplace: Main findings from the OECD AI surveys of employers and workers* presented by OECD IA [5]. The basis for the design of our questions was to collect the information necessary for the study in an objective manner, using simple, direct language and minimising bias. Further details can be found in the methodology section.

The web application to access the questionnaire was developed by IIA - LNCC. The data collected was handled in accordance with the Brazilian General Data Protection Law (GDPL)[2]. Further details can be found in the methodology section.

As Brazil is a continental country with regional specificities, we sought partnerships to maximize the scope of the study and thus capture the wide variety of responses that reflect the needs and aspirations of the respondents. However, this first evaluation took on greater proportions in the southeast region. According to the IBGE [14], the cumulative annual gross domestic product (GDP) in June 2024 grew by 2.5% in Brazil, mainly driven by the southeast region, which has the largest number of industrial and commercial activities and accounts for more than 50% of the country's total GDP. In this sense, it is worth mentioning that we sought several partnerships and it was essential to count on unions, associations and confederations to reach the target audience. These organizations represent the classes, synthesizing the needs of the group and intervening with governments to promote improvements and equality. They have direct contact with companies and workers and acted as intermediaries so that the survey could reach its target audience. They also contributed by reporting their experiences with applying other similar types of online surveys on the members of their groups and indicated some specific language expressions to make the understanding of the questions more appropriate. Our partners are Union of Data Processing Companies - SEPRO-SP/CNS (15), National Confederation of Commerce - Fecomércio-RJ/CNC (16), National Confederation of Journalists - FENAJ (17), National Association of Newspapers - ANJ (18), Association of Digital Journalism - Ajor (19).

The survey applied to the employee and the employer are presented in the [Methodological Annex](#).

## 3. Methodology

The research began with a search for technical material on the relationship between the tasks performed by AI and human cognitive abilities such as communication, manual dexterity, reasoning, learning and others. This understanding was relevant to the design of the questions that make up the survey.

---

[2] https://www.planalto.gov.br/ccivil_03/_Ato2015-2018/2018/Lei/L13709.htm

The methodology approach is based on an exploratory analysis with qualitative sampling. This research format brings benefits in relation to the bias of the researcher's influence and also minimizes the risk of distortion in misunderstanding when compared to interviews. The strategy on how to formulate and to apply the survey was discussed extensively with the Brazilian Institute of Geography and Statistics (IBGE). To minimize bias, the questionnaires were evaluated by a quality-evaluation expert after completion and underwent some major adjustments.

An important point that was observed from the start of development was the time taken to complete the form. The number of questions was limited to 30, most of them included alternative answers, aiming to reduce the response time. Depending on the needs of the question, open-ended responses were provided to capture individual perceptions specific to the topic.

Concerning information security, we decided to develop our application instead of using ready-made available form tools from large software companies. As of our observations, the tools we could identify on the market do not let users manage the security level in the lower layers. As a result, they introduce uncertainty about the security of access to the information captured and vulnerabilities that can be exploited in undesired ways. The IIA developed the application using a specific framework and database system. Security and privacy were improved, and scalability for concurrent access was ensured.

A third important point was to pay full attention to the guidelines of the General Data Protection Act - LGPD. In order to meet the requirements related to sensitive personal data, we involved the relevant sectors of the LNCC and the *National Research Network* (RNP)[3]. Among the legal requirements implemented are the transparency of information, validated by informative texts, the need for respondents' consent, encryption and anonymisation procedures applied to data during and after collection, and general rules for international data sharing.

We are committed to ethics and responsibility in the use of data, so the option of owning the application allowed IIA to fully manage information security, maintaining the integrity of the data in the database (no losses, no changes, no additions, no external leaks), that is, to guarantee secure data throughout its cycle, from the moment it is generated, collected, transmitted, stored, analysed, shared and disseminated.

Macro Steps:

1. Carrying out applied research to evaluate pre-existing methodologies;
2. Construction of questionnaires: Employee and Employer;
3. Validation of questions and adjustments in relation to the study objective;
4. Construction of the database schema and modeling;
5. Construction of the virtual hosting and storage environment;
6. Validation and security tests of the virtual environment;
8. Validation tests of forms and verification of successful data capture and storage;
9. External opening to capture responses and dissemination with the help of supporters;
10. Data analysis and final report [4].

The survey was conducted between the 5th of September and the 15th of October. To reach the target audience, the support of partners from unions, associations and confederations was essential. These organizations represent the classes, synthesizing the needs of the group and intervening with governments to promote improvements and equality. They have direct contact with companies and workers and acted as intermediaries so that the survey could reach its target audience. They also contributed by reporting their experiences with applying other similar types of online surveys on the members of their groups and indicated some specific language expressions to make the understanding of the questions more appropriate. Our partners are Union of Data Processing Companies - SEPRO-SP/CNS (16), National Confederation of Commerce - Fecomércio-RJ/CNC (15), National Confederation of Journalists - FENAJ (17), National Association of Newspapers - ANJ (18), Association of Digital Journalism - Ajor (19).

---

[3] www.rnp.br
[4] Further details are described in the Annex regarding the technical methodology.

The survey was promoted directly, through email marketing to the list of registered participants in each partner association, and through websites and social networks such as Linked-in, Facebook and Instagram. Reminder emails to increase participation were resent 15 days after the survey began, and after 30 days the deadline was extended with new updates and notifications through all channels. Participation was voluntary and access was facilitated by a link or QR code accessible from any electronic device.

For the statistical design of the research, the reference population was subdivided according to the size of our partners. We estimate that we will reach approximately 10% of the 75.000 active member companies that employ at least one of the following sectors: software development, customer service and graphic design. However, it was not possible to estimate the number of employees. For the journalism sector, we estimate about 40.000 associated journalists, although not all of them are active, and about 249 employers in journalism and content production in print or digital form. For the distribution and retail sector, we do not have an estimate of the number of employees. Given the uncertainty of the expected sample size, we aim for greater statistical precision with wider confidence intervals. However, this condition is not a limiting factor for stopping the study and can be improved in the future. The total mass of formal employees for each of the sectors cannot be used as a reference, as we don't have the full reach yet.

The survey applied to the employee and the employer are presented in the Appendix I.

### 3.1. Preliminary Results

The information extracted from the data in this survey is preliminary and reflects the current status as from the perspective of employers and employees in the target professional occupations reached through our partners. We recognize that the survey has some limitations from a sampling point of view, as the reachable population was biased by the partners that supported the survey. The majority of the answers came from the southeast region of Brazil. In addition, the total number of participants is still low, considering the target spectrum. However, the findings obtained may raise generalizable questions that help in the preparation of measures to deal with technological changes.

The modular form of the questionnaire allowed for greater data management. It is possible to store the data partially, with each section completed and if they drop out at any time before the end of the stage the data is discarded. We therefore had n=95 participants who completed the first section. Dropping out of the questionnaire is expected and foreseen. The total number of respondents at the end of the third module was n=69. The dropout rate after participation was 18% for employees and 32% for employers.

### 3.2. General trends

The number of respondents by "Employees" and "Employers" segmentation who answered the first section of the form is not the same as the number who remained until the end. But the proportion remained practically the same, Figure 3.

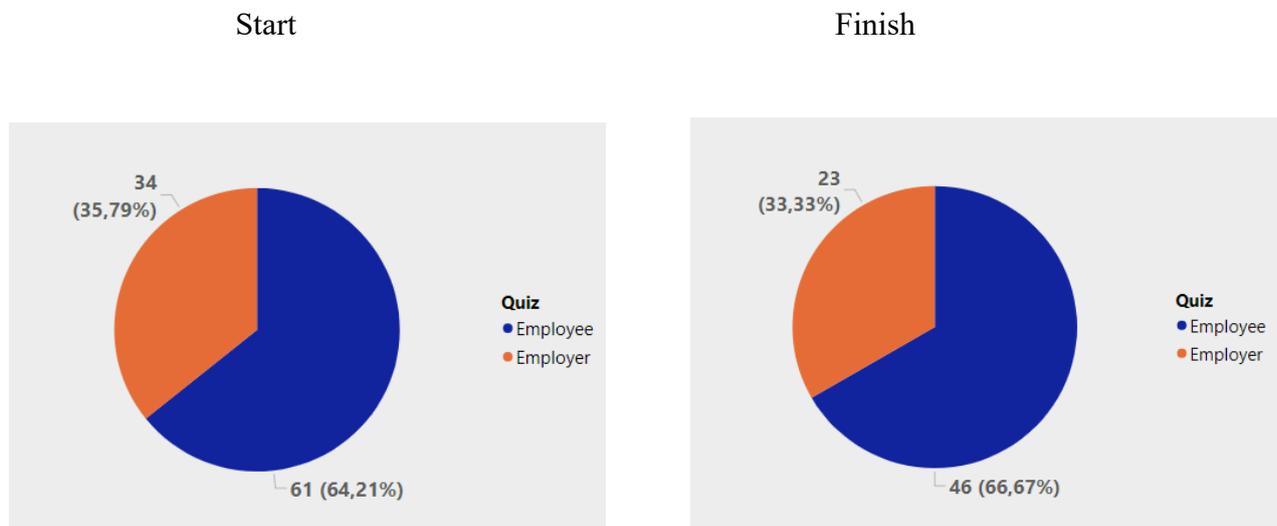

Figure 3. 'Start' shows the number of respondents who started the survey and 'Finish' the number who remained until the end of the survey.

The "Employers" survey reached six brazilian's states, located in the South (Rio Grande do Sul e Santa Catarina), Southeast (São Paulo, Rio de Janeiro e Espírito Santo) and Northeast (Ceará). The city of São Paulo accounts for almost 62% of the results, the highest among all cities, Figure 4.

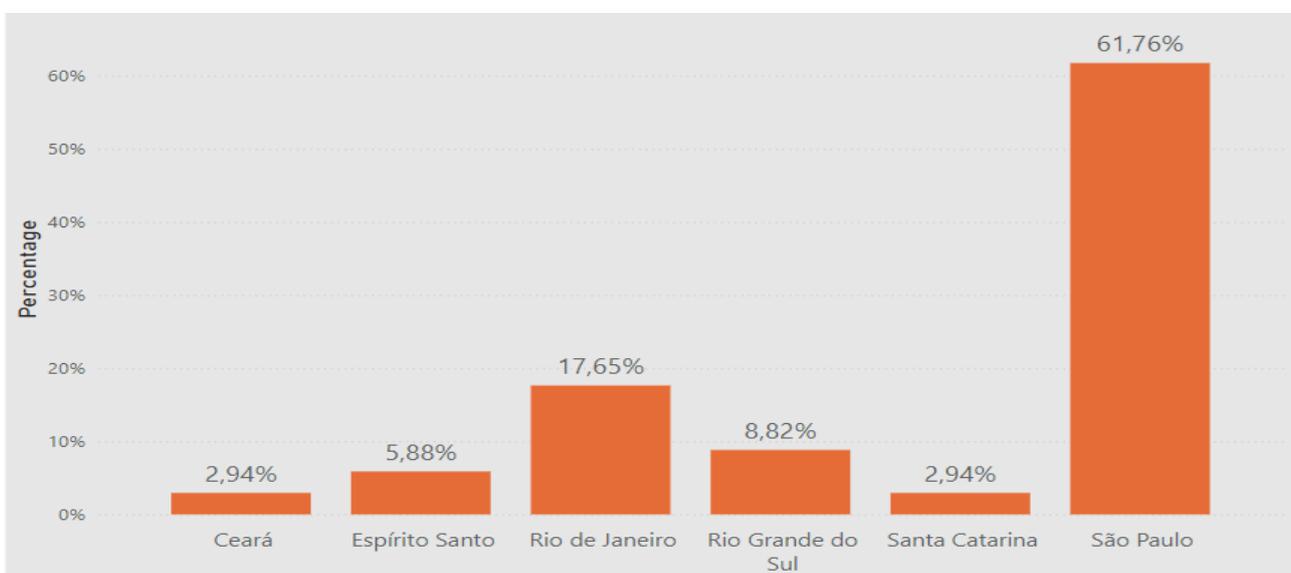

Figure 4. States with respondents from the "Employers" survey (sample size=34).

According to the Brazilian Micro and Small Business Support Service (SEBRAE Comércio Serviços) [20], companies are classified by the number of employees: *Micro-enterprise* up to 9 employees; *Small-enterprise* from 10 to 50; *Medium-enterprise* from 51 to 100, and *Large-enterprise* above 100. 79% of the Brazilian registered companies are classified into the micro or small company groups. In this survey, we had 34 companies participated. Figure 5 depicts the classification of these companies following the SEBRAE's distribution.

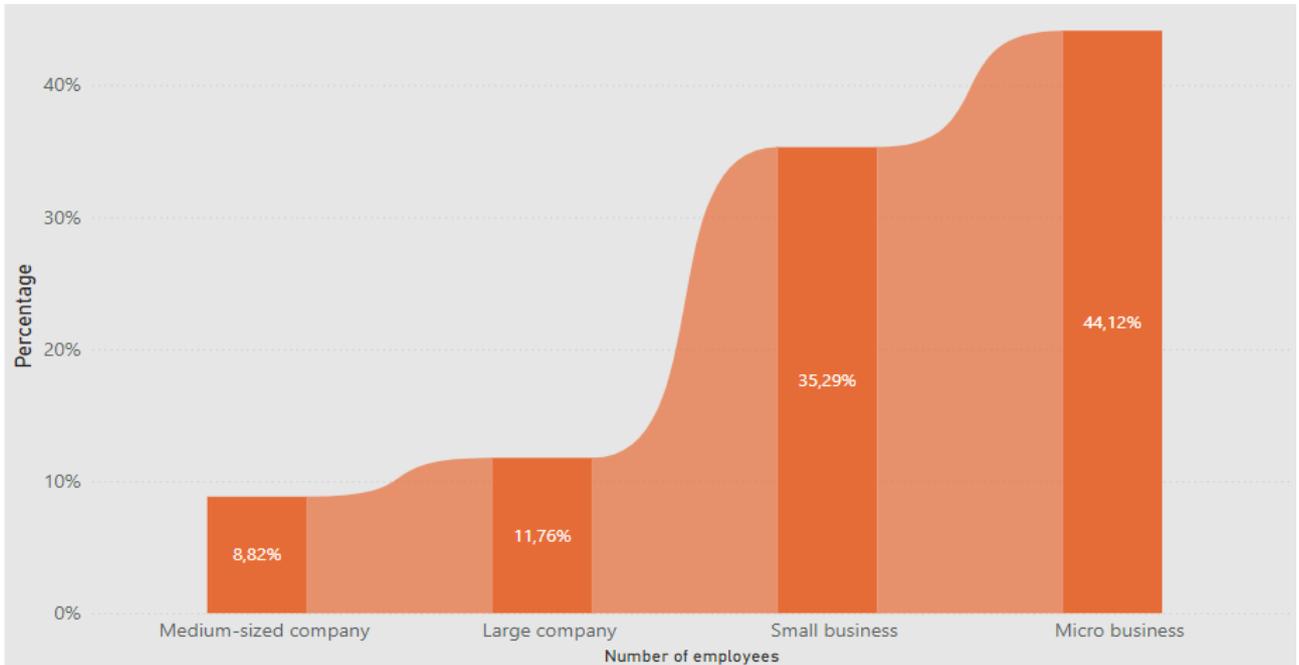

Figure 5. Respondents' companies classification (sample size=34).

The "Employees" survey covered 31 municipalities located in the South, Southeast, Central-West, Northeast and North regions, Figure 6.

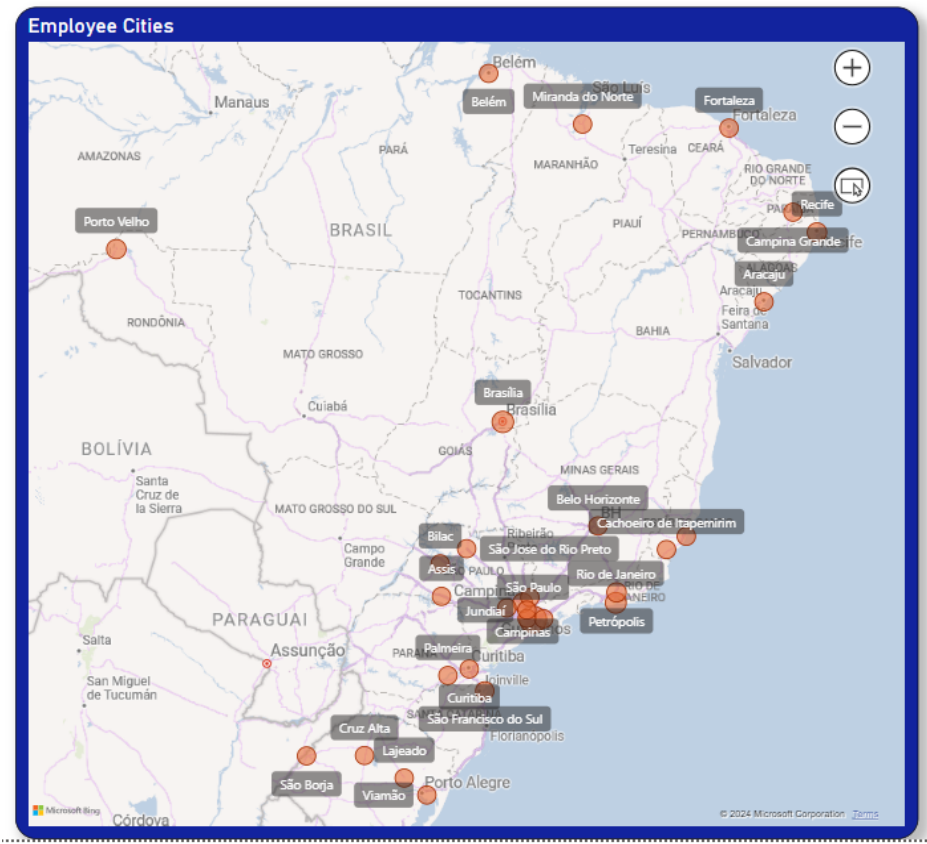

Figure 6. Cities with respondents for the "Employees" survey.

The segmentation of "Employees" by professional occupation shows that *Journalism* or *Content Production* accounted for the highest percentage of participation, at 50%. *Sales* accounted for the lowest percentage, with 4%, Figure 7.

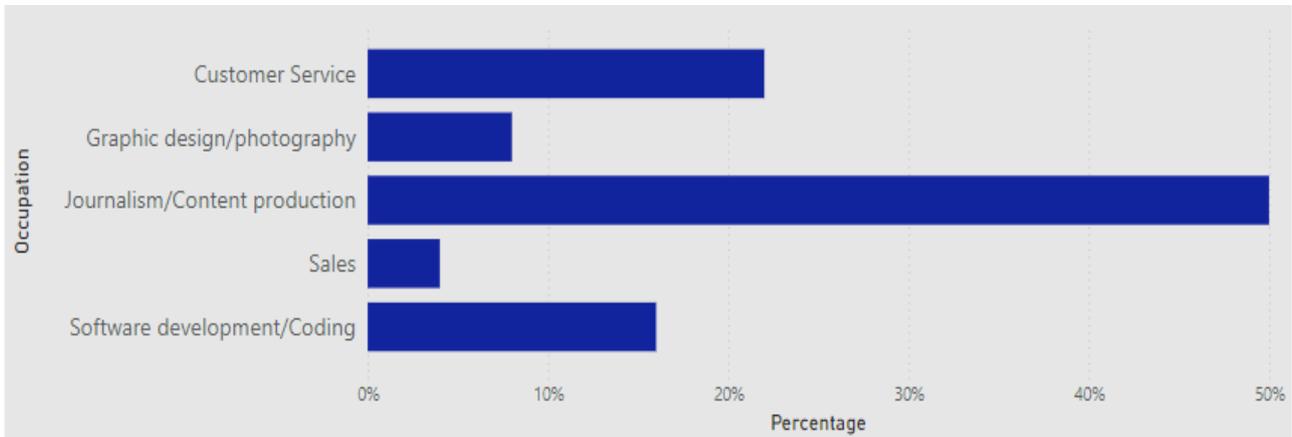

Figure 7. Categorization by "Employees" professional occupations (sample size=50).

It is important to note that the legal representatives: managers, partners or directors of companies who responded to the survey as the role of "Employers", checked the existing professional occupations within the corporation they represent. This was a multi-format implementation, unlike the employee form, whose response is individual and personalized, so it was acceptable to mark multiple sectors, Figure 8.

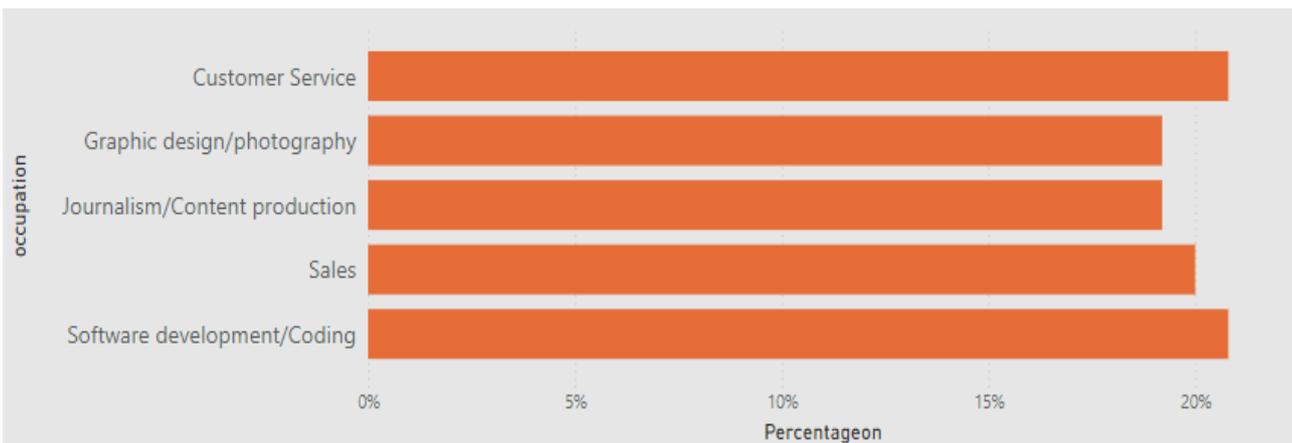

Figure 8. Categorization by "Employers" professional occupation (sample size=125).

From the employers answers, it could be inferred that their companies include activities for all professional occupations of interest. This doesn't seem to match the reality of the data, especially since most of the participating companies are *micro* or *small-companies*, according to the results already presented. Another argument that corroborates our identification that there was a failure in capturing the data for this question is from the professional occupation point of view, once some of which are quite distinct and therefore minimize the chance of the same *micro-small* company holding departments or positions for those occupations. The degree of uncertainty regarding this result is high and therefore this profile will not be considered in the exploratory assessments.

## 3.3. Employee group

This section analyzes the employee group that answered the survey through the lenses of different dimensions.

**Age**
The age range spread between 20 to 72 years old. We got the participation across many generations, which was great to raise different points of view about the new technology, Figure 9.

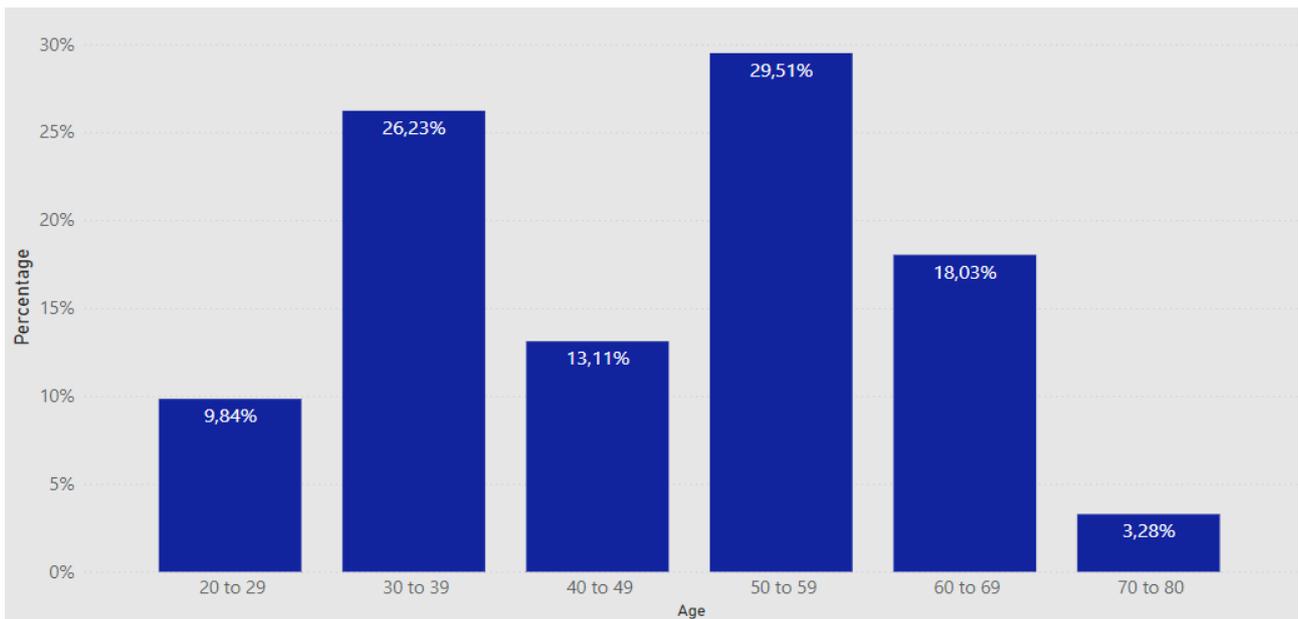

Figure 9. Distribution of the Employee group by age (sample size=50).

**Educational Level**
The level of education observed starts from a minimum of complete secondary education with 5% of the responses. The majority of respondents, 59%, have a Postgraduate degree or similar post-undergraduate degree, Figure 10.

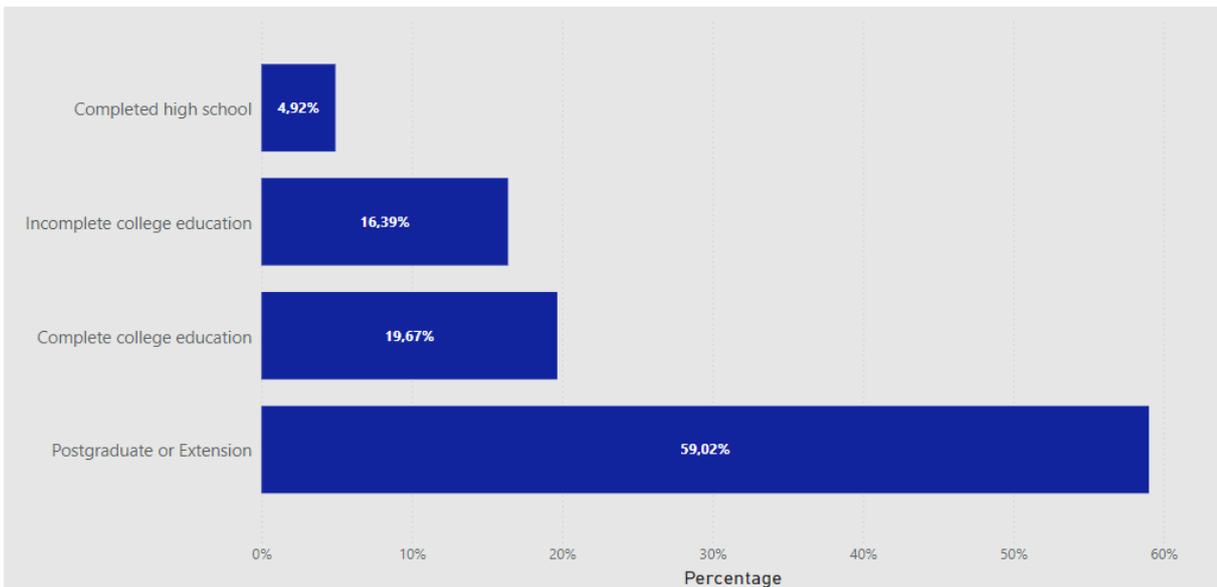
Figure 10. Distribution of the Employee group by education level (sample size=50).

**Salary Range**
The salary scale is well balanced for all ranges, with the following percentages: above 10K 26%; from 6.5K to 10K 14%; from 4K to 6.5K 20%; from 2.8K to 4K 18%; from 1.4K to 2.8K 22%, Figure 11.

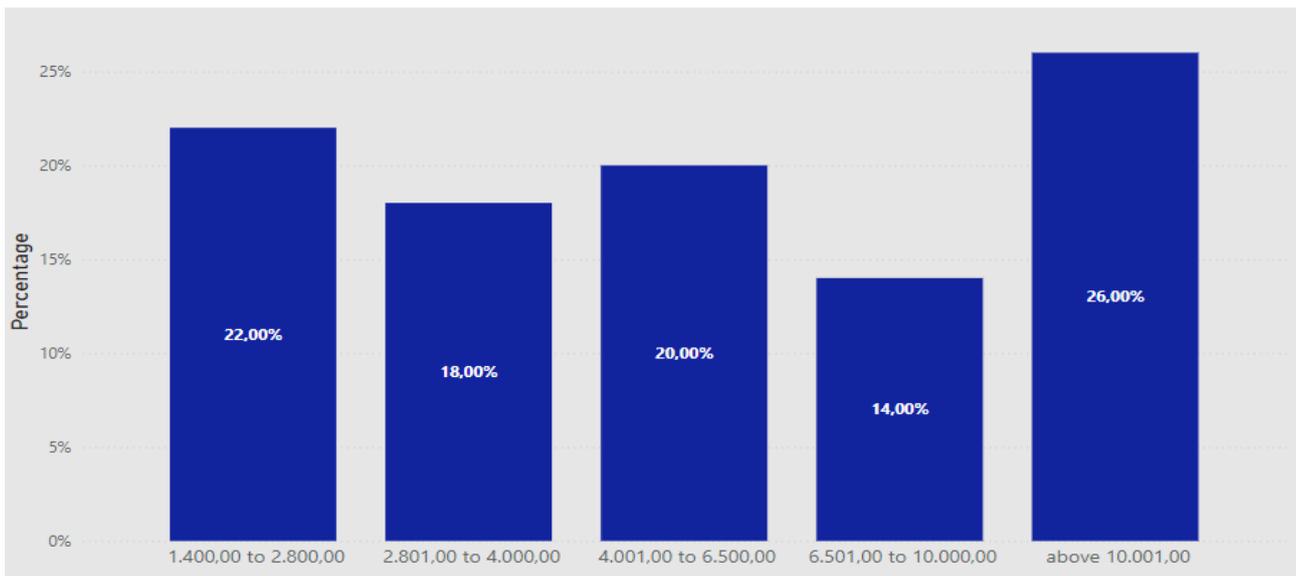
Figure 11. Distribution of the Employee group by Salary range (sample size=50).

**Type of work contract**
The majority of employees work under the contract type known as the Consolidation of Labor Laws "CLT", which represents 40 hours per week and all regular labor benefits, with 66%. In Brazil there is a variation of contract types departing from the traditional regular 40 hours with full benefits. We aggregated those different modalities into a single class. Among those classes, the most representative covers 6% of the group and corresponds to people working under a contract between the employer company and her individual company, Figure 12.

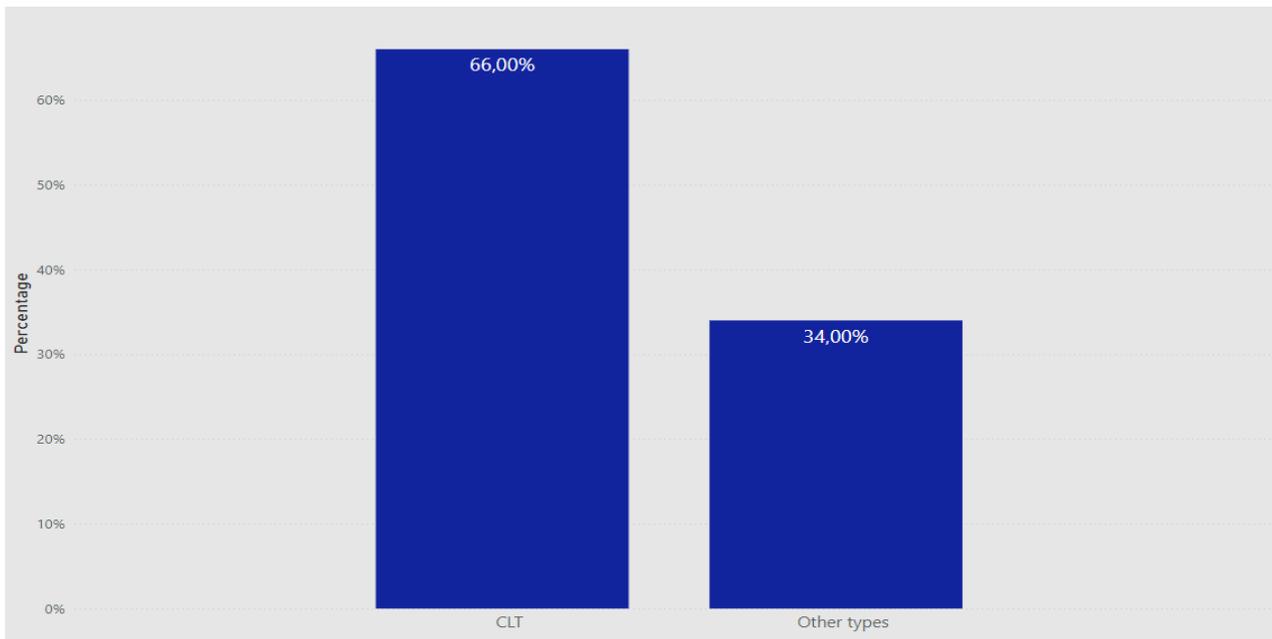
Figure 12. Distribution of the Employee group by Type of work contract (sample size=50).

**Mode of work**
The mode of work is the form of location from which an employee performs their duties —remote, hybrid, or onsite, with onsite, with 60%, followed by hybrid, with 34%,and lastly, with 6%, remote work, Figure 13.

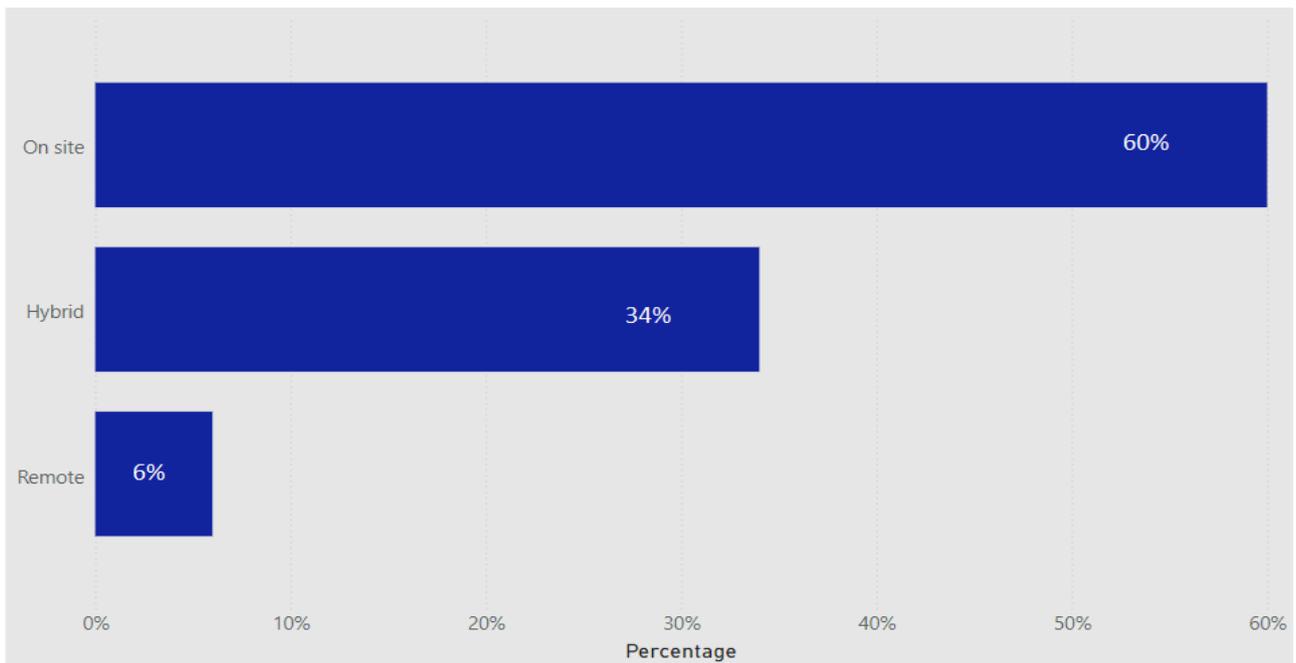
Figure 13. Distribution of Employees by work mode (sample size=50).

**Time in and out of work registered**
In Brazil, employees may be obliged to register the time they arrive and leave office. 60% of the respondents informed they fit into this category, Figure 14.

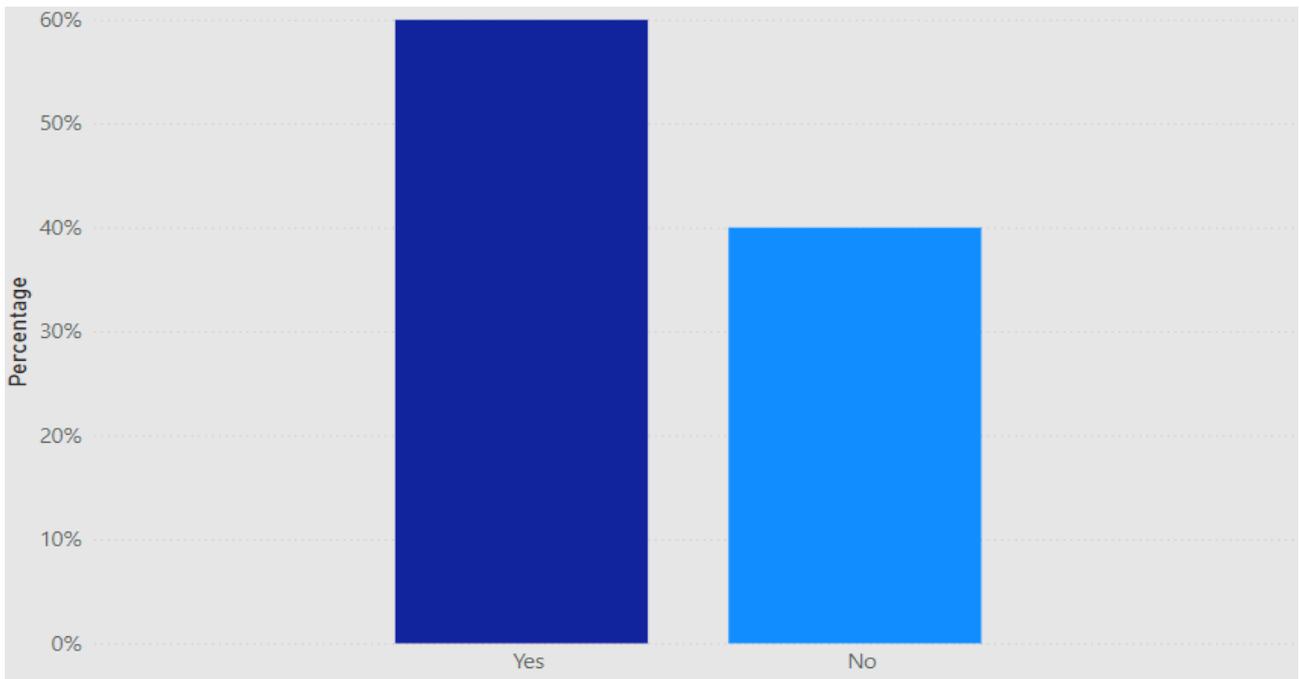
Figure 14. Percentage of respondents daily recording arrival and departure time (sample size=50).

**Sharing of work tasks**
38% of employees do not share their tasks with colleagues, they perform their functions individually, Figure 15.

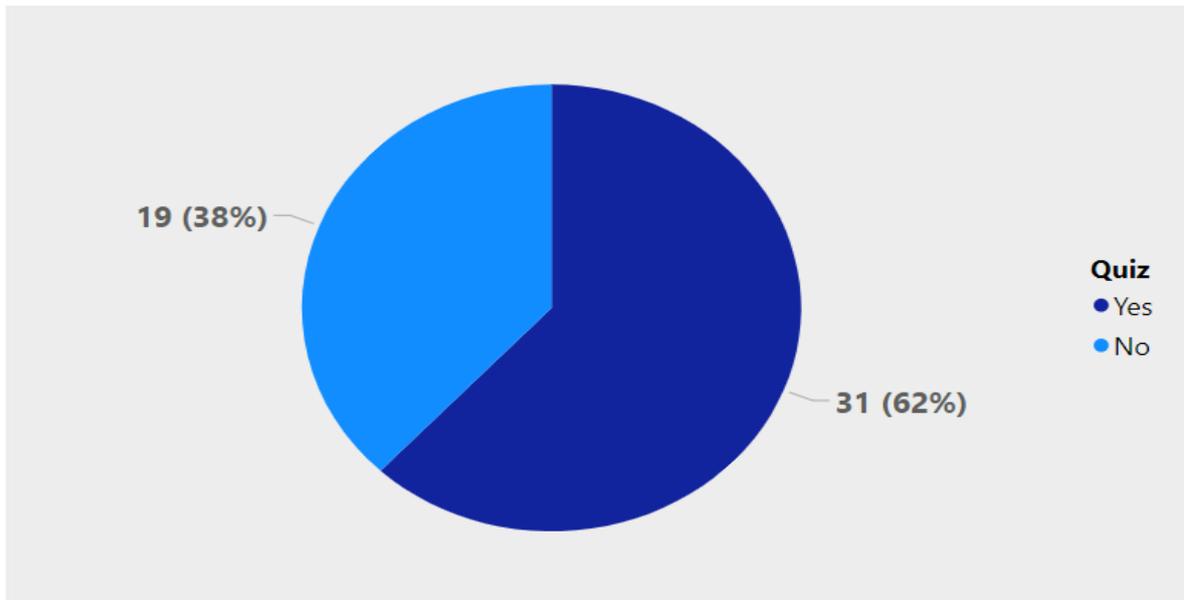
Figure 15. Distribution of employees by Collaborative work (sample size=50).

**Workload**
This dimension evaluates the employee's perception with respect to his/her workload.
78% of respondents consider their workload to be fair, Figure 16.

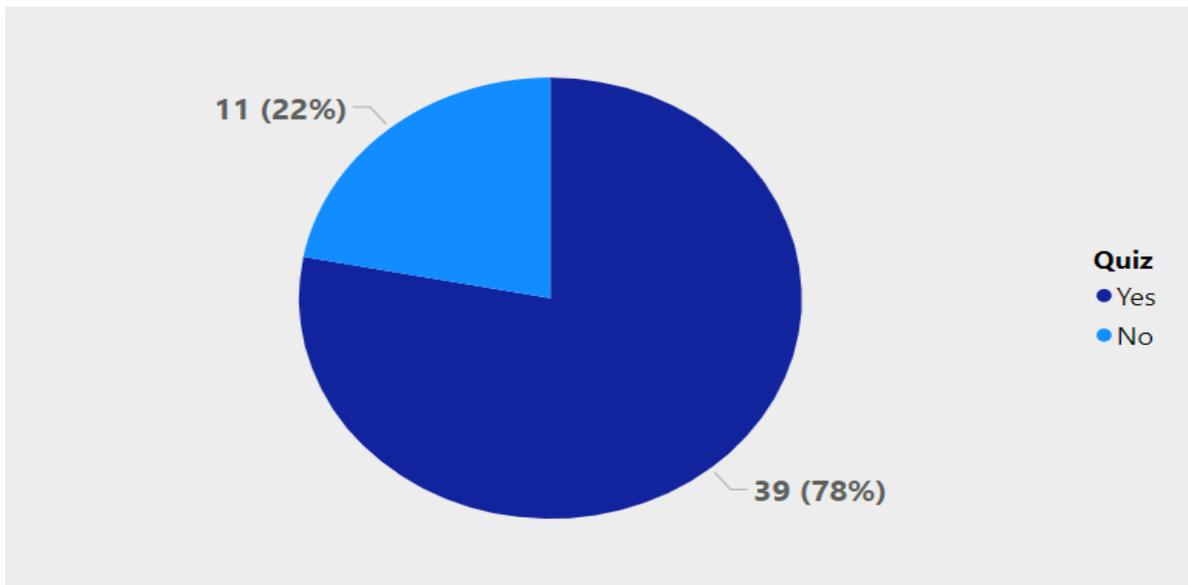

Figure 16. Distribution of Employees by their perception on their workload (sample size=50).

**Benefits**

This dimension enables employees to identify the benefits they receive from their employment contract. Each employee may indicate more than one benefit, Figure 17.

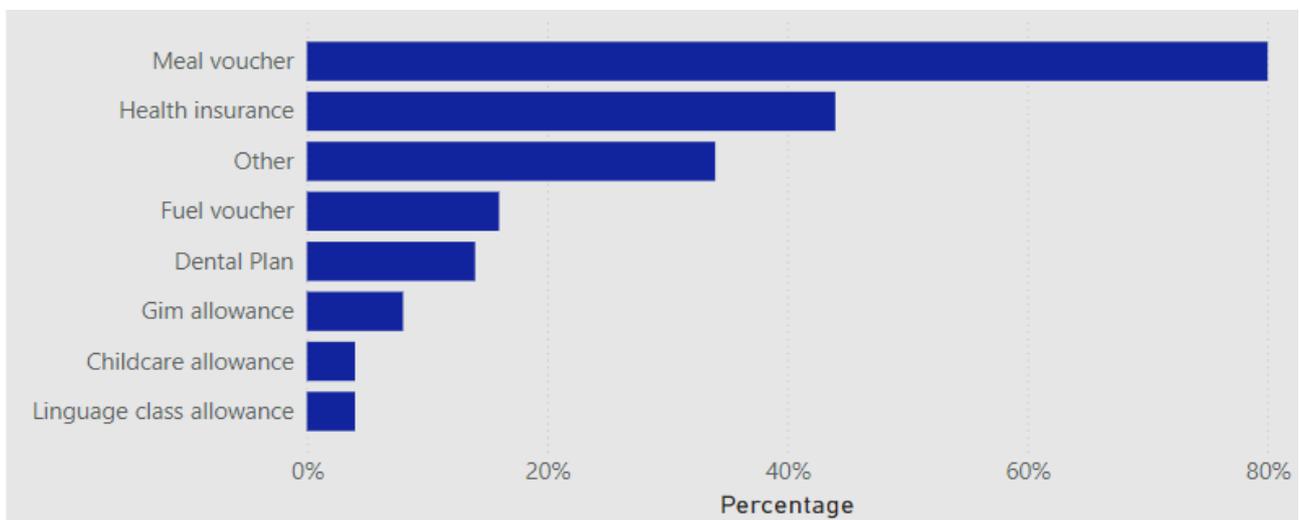

Figure 17. Distribution of benefits among respondents Employees. (sample size=102)

This question offered an open entry field so that employees could enter other benefits. Employees registered *health-related* benefits such as: *health assistance* or *health insurance*, and *commuter benefits*. Few people don't receive any benefits due to the type of contract they have. The other benefits cited are related to extra payments linked to a leadership role, the achievement of personal or company goals.

**Additional Compensation or performance bonus**

According to the survey, 60% of participants stated they receive no additional compensation., but assert that it would be motivating for their daily work if they could receive it, Figure 18.

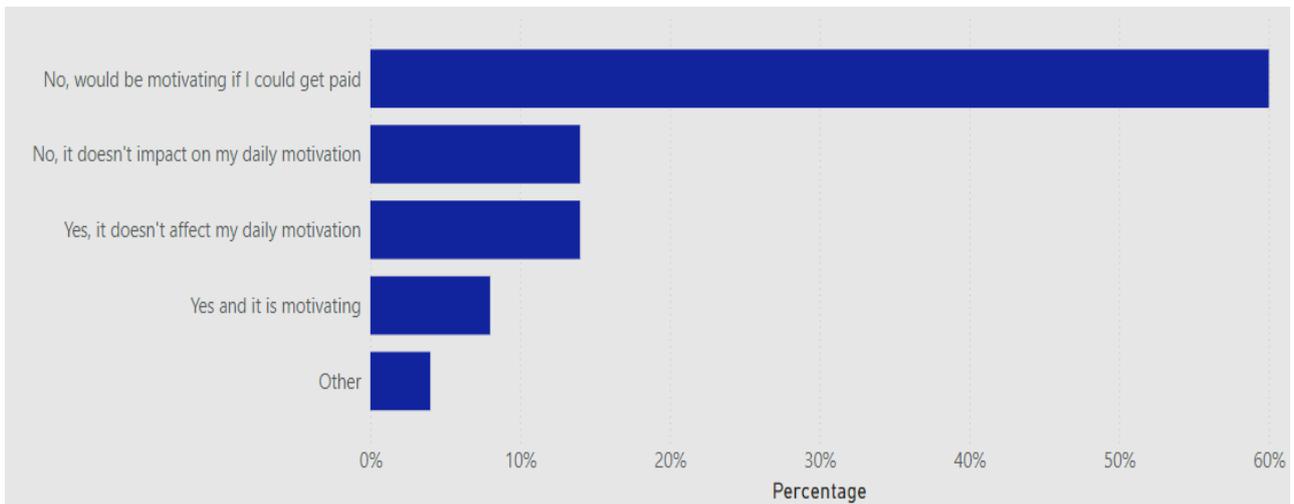

Figure 18. Distribution of perception of additional compensation or performance bonus. (sample size=50).

**Workplace conditions**

Regarding the conditions of the workplace, the workers expressed their feelings about the physical space, ergonomics and technological issues made available for their daily work. The majority classify it as Good for all work conditions, Figure 19.

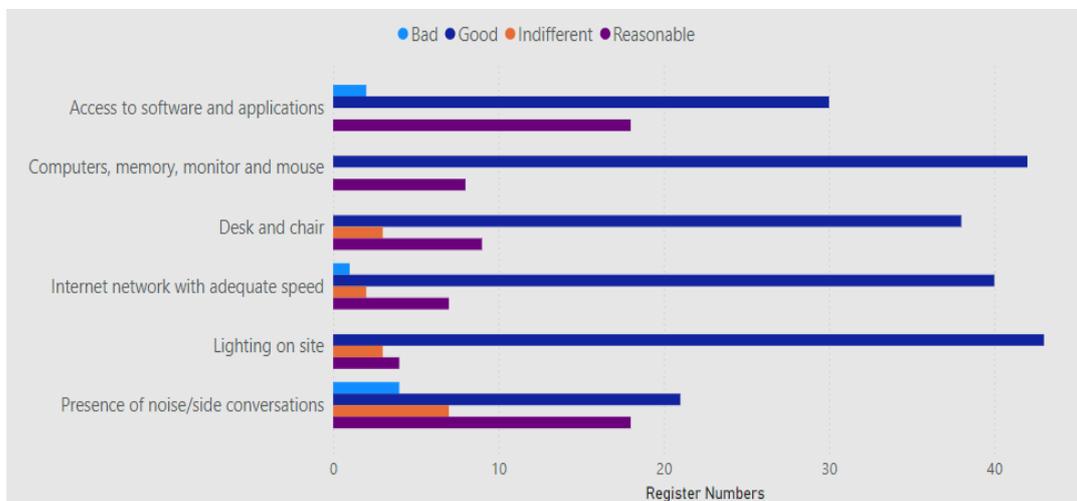

Figure 19. Distribution of perception of Workplace conditions.

Figure 20 presents a table to better understand the percentage of Employees perception.

| Office ammenities | Bad | God | Indifferent | Reasonable |
|---|---|---|---|---|
| Acess to software and applications | 4% | 60% | | 36% |
| Computers, memory, monitor and mouse | | 84% | | 16% |
| Desk and chair | | 76% | 6% | 18% |
| Internet network with adequate speed | 2% | 80% | 4% | 14% |
| Lighting on site | | 86% | 6% | 8% |
| Presence of noise/side conversations | 8% | 42% | 14% | 36% |

Figure 20. Distribution of the perception of Employees about office amenities.

This is a profile of the employees who accepted this form format and responded voluntarily, contributing to obtaining this specific representation. Several categories of employment were detected among the respondents. The highest percentage, 66%, benefits from a formal employment contract. In Brazil, this work contract modality is named after the *Consolidation of Labor Laws* (CLT), which refers to the law that regulates formal employment. Other modalities of contracts identified in the survey are government employees, self-employed workers, trainees and others. It can be seen that the educational level of respondents is at least *higher education* for professionals in the sectors surveyed. This was to be expected as some of the professional occupations surveyed require a certain degree of specific skills, which are usually acquired in a formal college course. Respondent ages covered a wide spectrum, from 20 to 72 years old, which provides an interesting lens on the impact of technology in different age groups. Salary distribution shows a certain balance of representation. The lower limit approximates to the minimum legal wage in Brazil, R$ 1,412.00, and the upper limit was R$ 10,000.00, which is an expected value for more experienced workers. Thus, the perception is of an employee sample from a more intellectual part of the population, covering all career stages, with different types of working contracts and having access to benefits and work amenities, specially with good internet, which is relevant for this survey.

### 3.3.1. GenAI Trends

The third segmentation of the survey focuses directly on Generative Artificial Intelligence (GenAI). The questions addressed in this stage correlate with the previous ones to identify the changes experienced by the participants. The first evaluations refer to the employee questionnaire, followed by the evaluations for the employer.

### 3.3.2. Employee Perception

This section discusses the answers given to each of the questions raised in the survey section concerning the perception of the employee with respect to the impact of Generative AI tools in their professional activities. Remember that we are only considering professional occupations as listed in Table 1. The questions are presented in the same way and in the same order as the application questionnaire.

**Which sentence best describes your interaction with GenAI tools? Figure 21.**

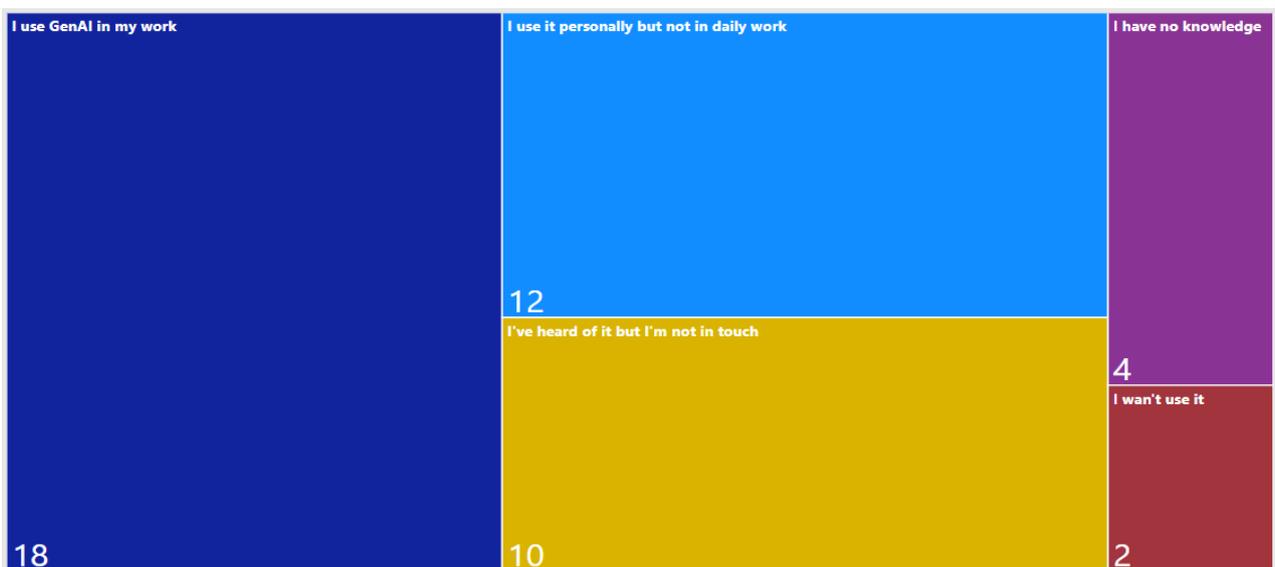

Figure 21. The evaluation of Interaction with GenAI tools (sample size = 46).

For 39% of the employees who responded to this section of the survey, GenAI is already part of their daily work routine. 26% use GenAI personally but have not yet applied it in their work. 22% heard of GenAI but have not had any contact with any tool. 9% say they have no knowledge about the tools, and 4% say they have heard of the technology but have no interest in using it.

**Do you believe that you have the qualification requirements to be able to use GenAI resources as a tool to help you with the tasks you need to perform in your job? Figure 22.**

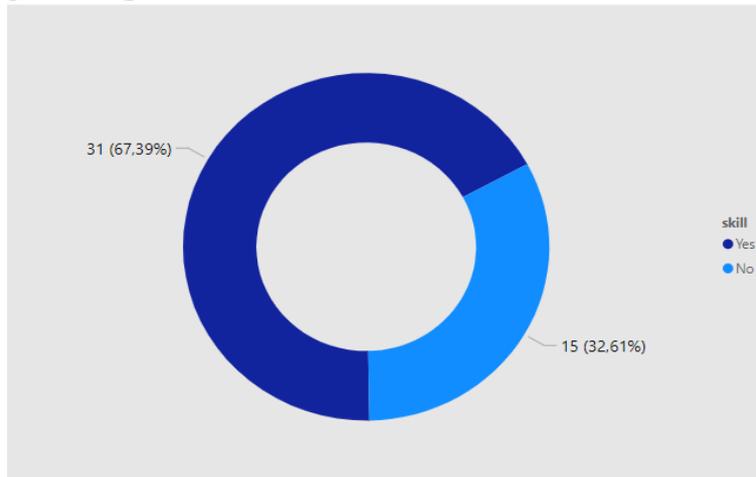

Figure 22. Qualification skills (sample size=46)

It can be seen that just over two thirds of employees assert that they have the necessary skills to use GenAI.

**Have you noticed that any intermediate leadership positions/decision-makers such as: managers, supervisors, coordinators, leaders have been impacted by the use of Generative AI? Figure 23.**

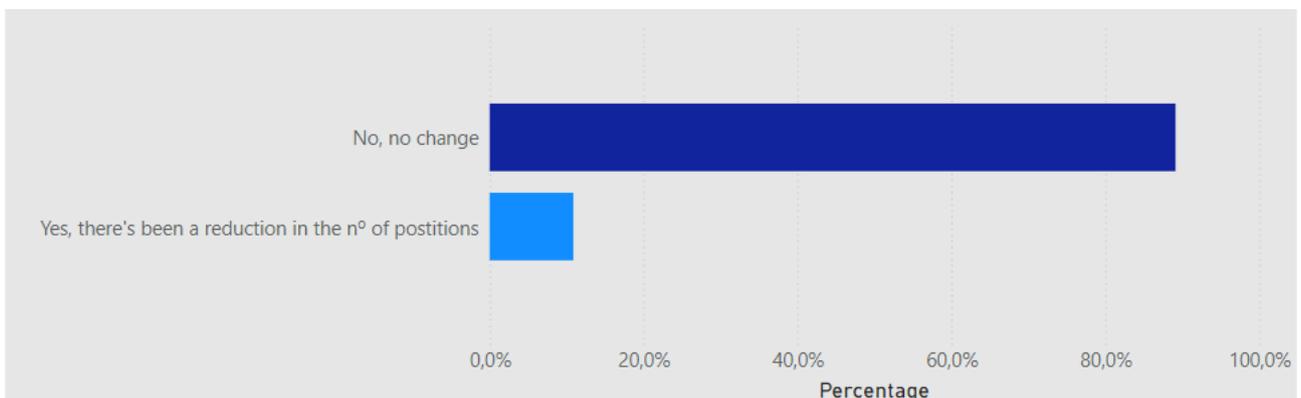

Figure 23. Impact in number of intermediate positions (sample size=46).

When asked whether or not intermediate leadership or decision-making positions, such as managers, supervisors and coordinators, will be replaced by GenAI, the majority of employees (89%) do not currently perceive this situation.

**In the last 6 months, have you used any GenAI tools for your activities, even in the testing phase? Figure 24.**

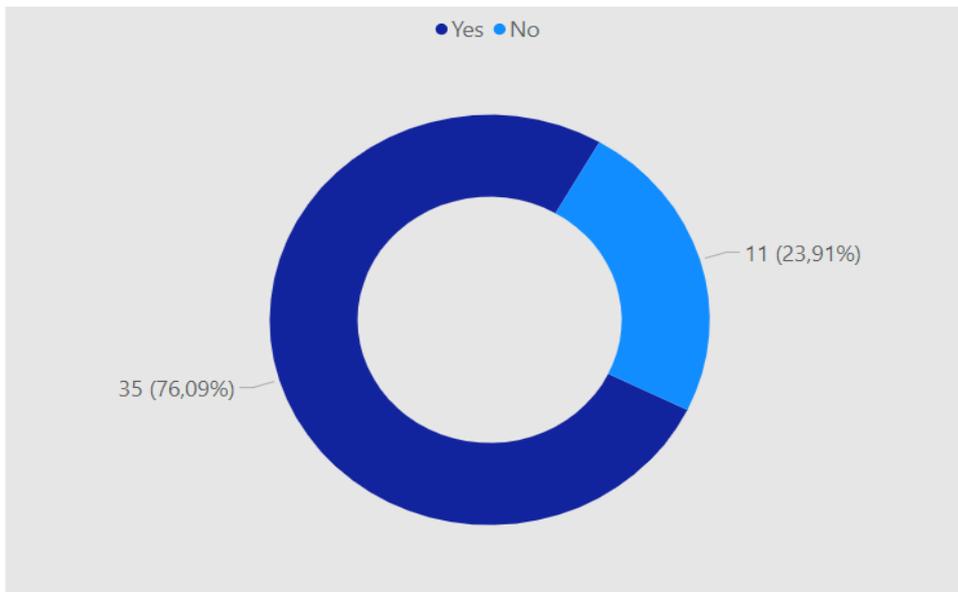

Figure 24. Use of GenAI tools (sample size=46).

24% have notuses GenAI tools in their work activity, not even in the testing phase. The other 76% have already made some use of the tools.

**What would be the barrier(s) to not adopting any AIGen tool? Figure 25.**

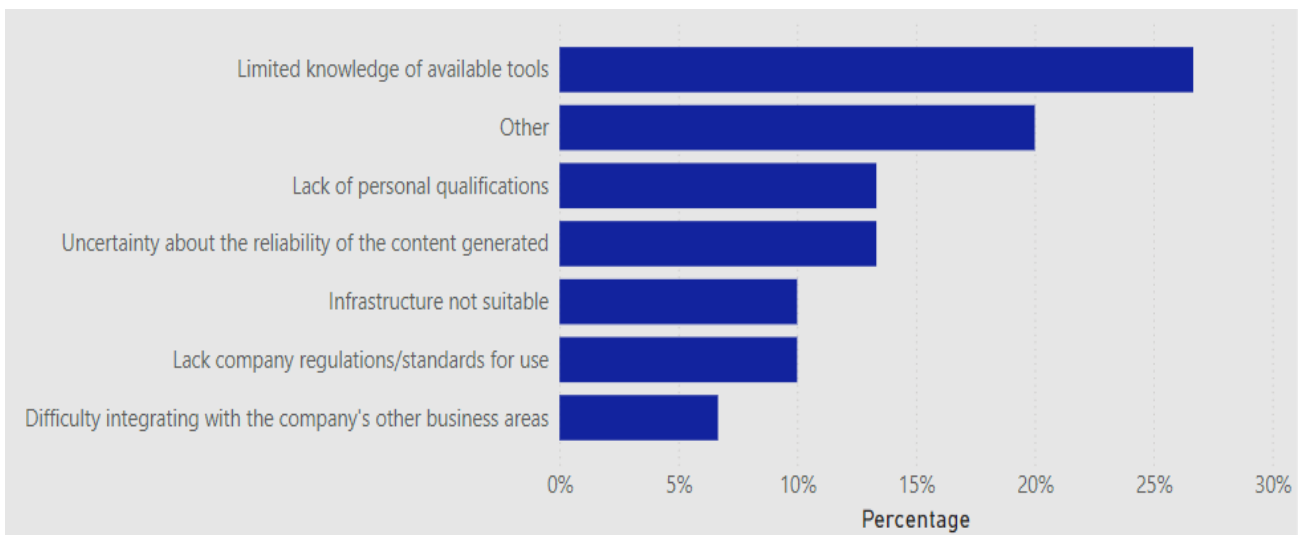

Figure 25. Barriers to not adopting the GenAI tools (sample size=30).

This question is not mandatory and enables multiple selections as answers. The most important answer was *Limited knowledge of available tools*. A open entry field, not shown in Figure 25, highlights that some respondents have a feeling that the GenAi tools don't add value, but just get in the way of the time needed to carry out daily processes.

**If you answered "Yes", have you ever used a GenAI tool, please indicate which one(s). Figure 26.**

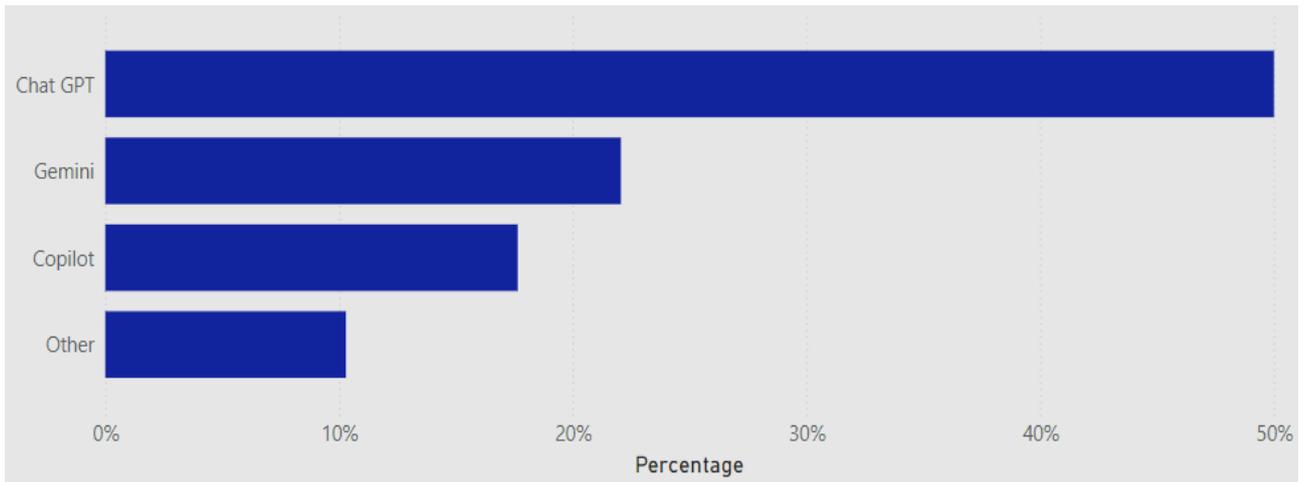

Figure 26. GenAI tools used (sample size=68).

Those who had used some type of GenAI tool in the last 6 months were asked to indicate which tools. This question accept multiple answers. The most frequently used tool was Open AI's GPT Chat, followed by Google's Gemini and Microsoft's Copilot. Other applications mentioned included Gamma, which is used to help create presentations and web pages, Midjourney and Leonardo.AI, which generate images from text descriptions, online meeting assistants, among others.

**Have you ever created a Generative AI tool of your own? Which segmented tasks. Figure 27.**

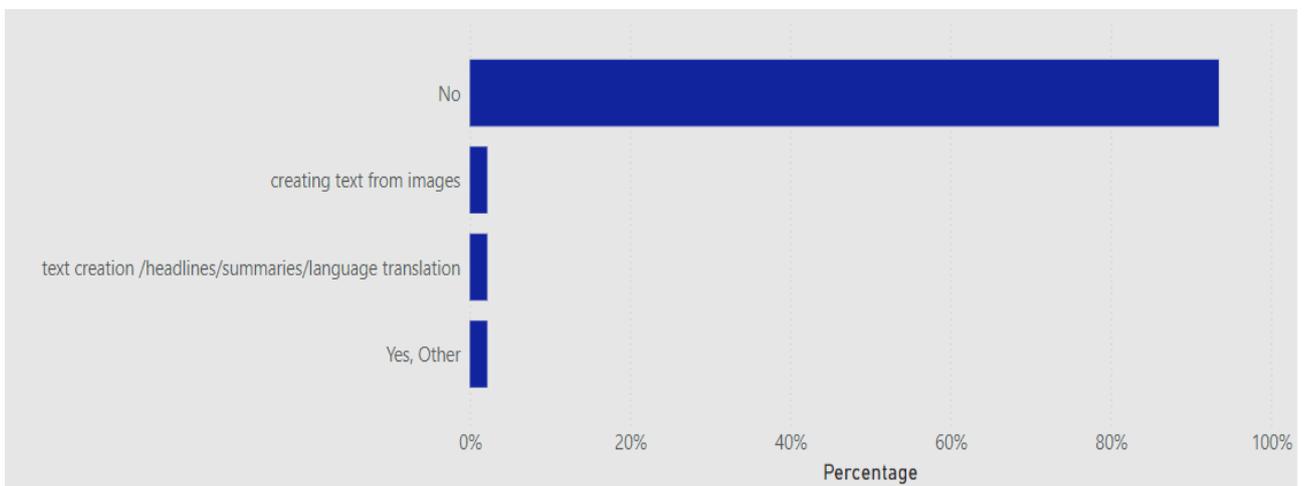

Figure 27. GenAI tools created by own (sample size=46)

94% of employees said they had not created any tools. The other 4% created tools associated with text generation from images and other sources. 2% created other data modalities. These responses reveal a very specific characteristic of the audience of coders and software developers.

**Have you observed any changes in your work as a result of using the Generative AI tool(s)? Figura 28.**

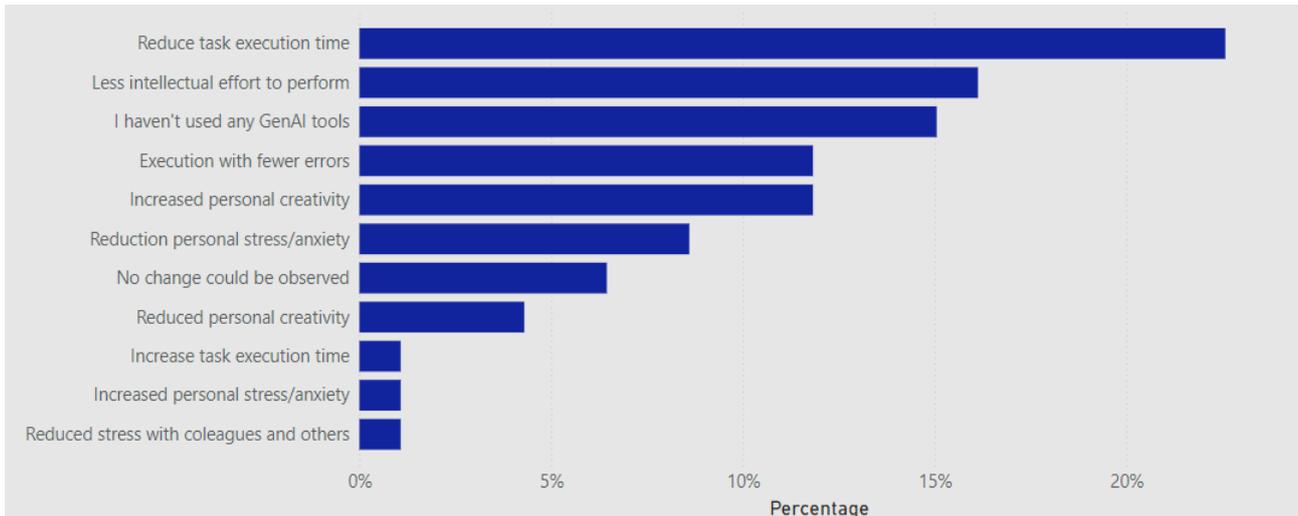

Figure 28. Change in work activities (sample size=93)

This question accept multiple answers. "Reducing the time needed to perform tasks" was the most cited change, followed by "Less intellectual effort needed to perform tasks".

**Did you receive any salary increase, additional benefit or notice any improvement in the working conditions offered by the company such as (internet network, computers, software, furniture and work environment) during or after the application of the Generative Artificial Intelligence tools? Figura 29.**

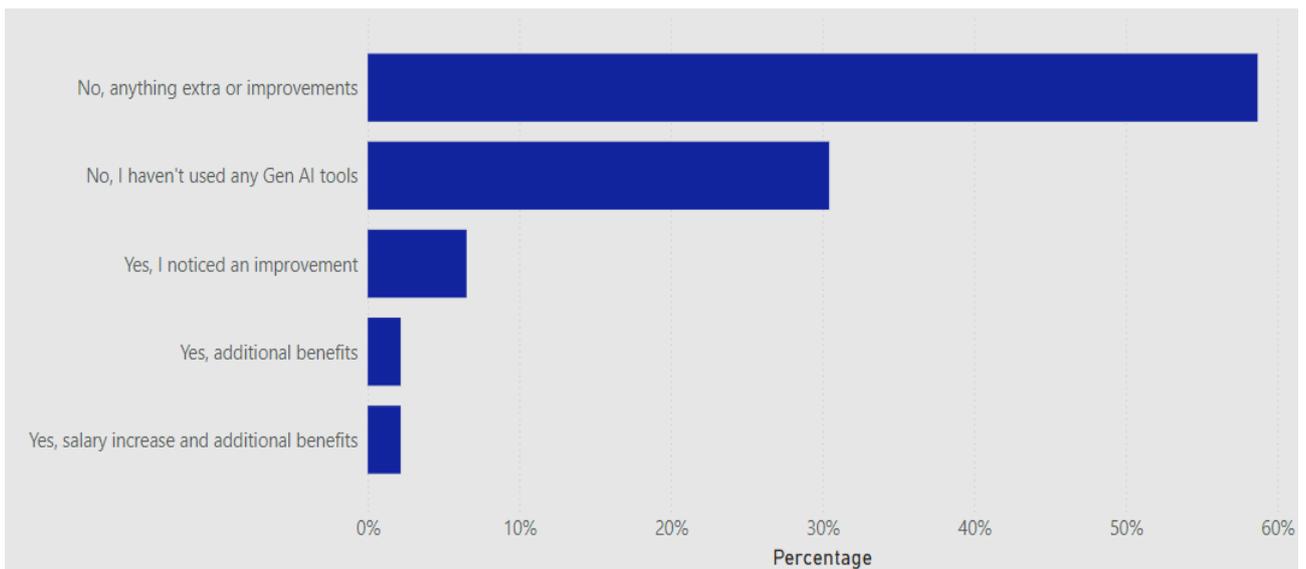

Figure 29. Report on salary increase or improved conditions. (sample size = 46).

59% of the respondents did not receive any extra remuneration or receive any additional benefits from using generative AI tools in their activities. 11% have seen some kind of improvement through extra payment or working conditions.

**Do you feel that your employer values an employee who is more knowledgeable about the use of Generative AI tools? Figure 30.**

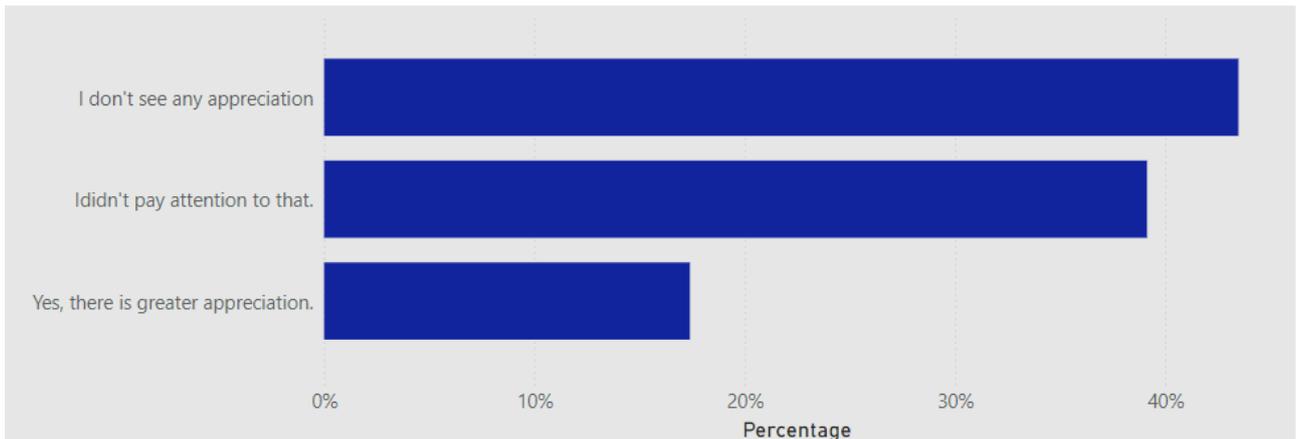
Figure 30. Knowledge valorization (sample size = 46).

39% are not concerned with the attention the employer dedicated to the adoption of GenAI tools, 45% don't see any appreciation from their employer regarding Gen AI knowledge, and only 16% perceive that their employers value their knowledge of new technologies more highly.

**Do you notice that your employer invests in training or lectures aimed at increasing theoretical or practical knowledge about the use of Artificial Intelligence? Figura 31**.

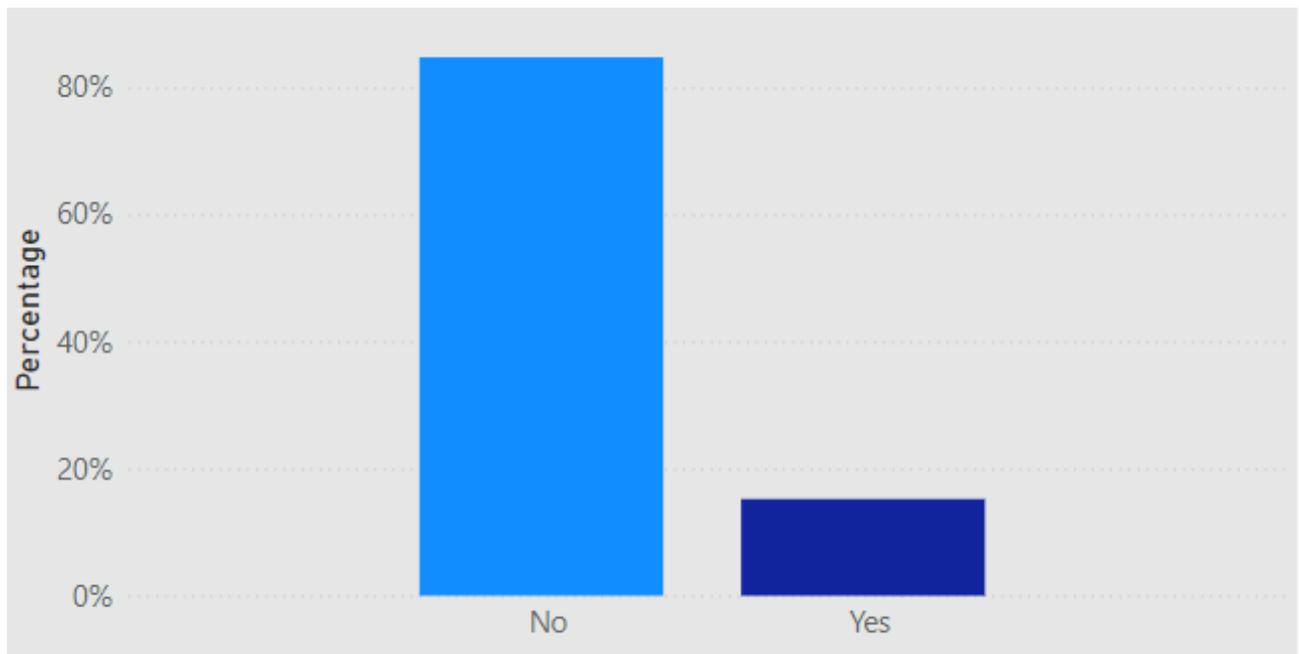
Figure 31. Perception of the investment in training about AI (sample size = 46)

85% do not see their employers investing in training to qualify their workforce specifically on GenAI.

**Do you learn how to use Generative AI tools on your own? Figure 32.**

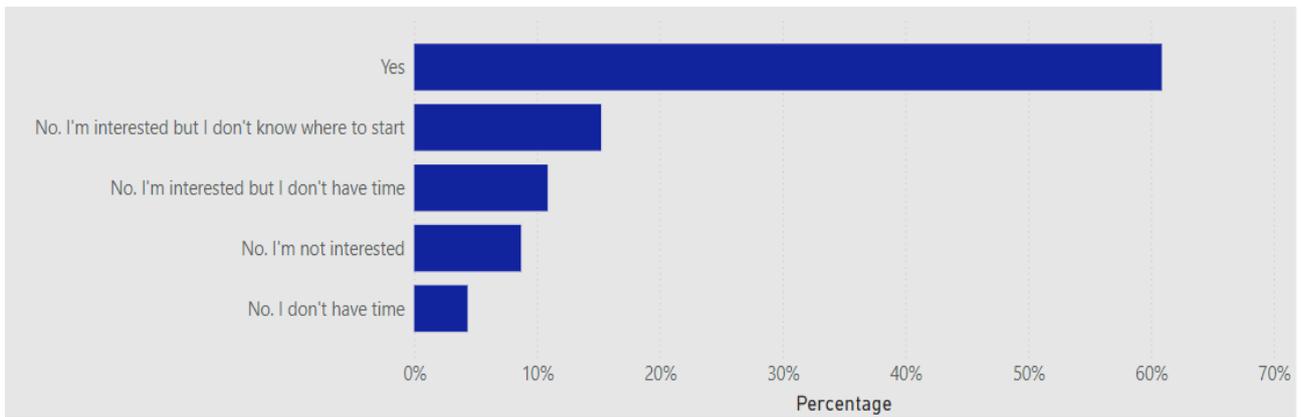

Figure 32. Learning about AI on your own (sample size = 46)

61% seek learning on their own mainly using the internet as a source. Another 16% have interest in learning about GenAI but don't know how to start. And 11% don't have interest in learning it.

**In the last 6 months, have you seen any of your colleagues dismissed from their jobs as a result of their tasks being replaced by Generative AI? Figure 33.**

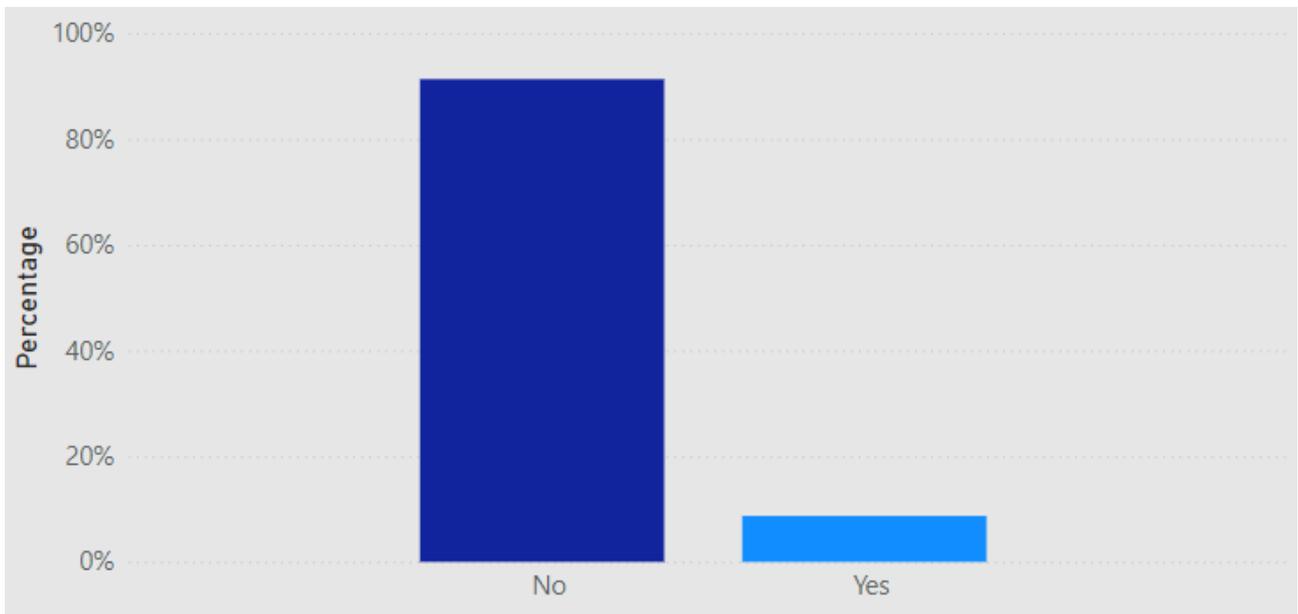

Figure 33. Experience of dismissed colleagues by GenAI substitution (sample size= 46).

9% observed colleagues dismissed as a result of their tasks being replaced by GenAI. The occupations of *content production* and *manager* were cited.

### 3.4. Employer group

In this section, we present and discuss the answers obtained from the Employers who responded to their perception about the impact of GenAI on their activities. The presentation will follow the order in which the questions appeared in the questionnaire.

**Which sentence best describes the company's vision regarding GenAI tools? Figure 34.**

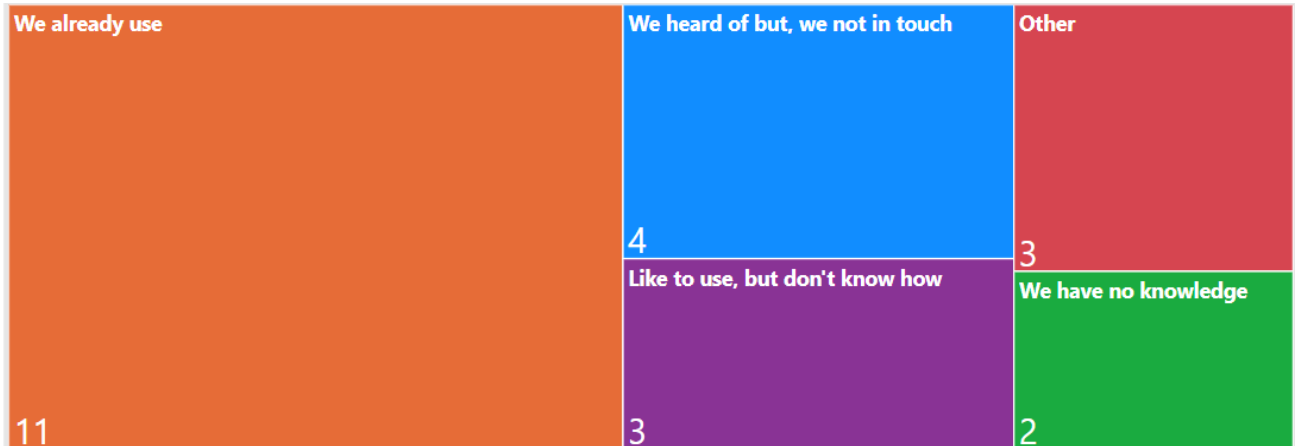

Figure 34. Perception of engagement with GenAI tools(sample size=23).

48% identify the phrase "I already use GenAI" to represent the company's vision. 17% have heard of it, but the company is not in contact with this technology. 13% would like to use it but don't know how to get started, while another 13% are starting some applications in the test phase or waiting for the technology to become more consolidated before joining it. 9% have no knowledge of the technology.

**In the last 6 months, have any of these sectors used a GenAI tool, even in the testing phase? Figure 35.**

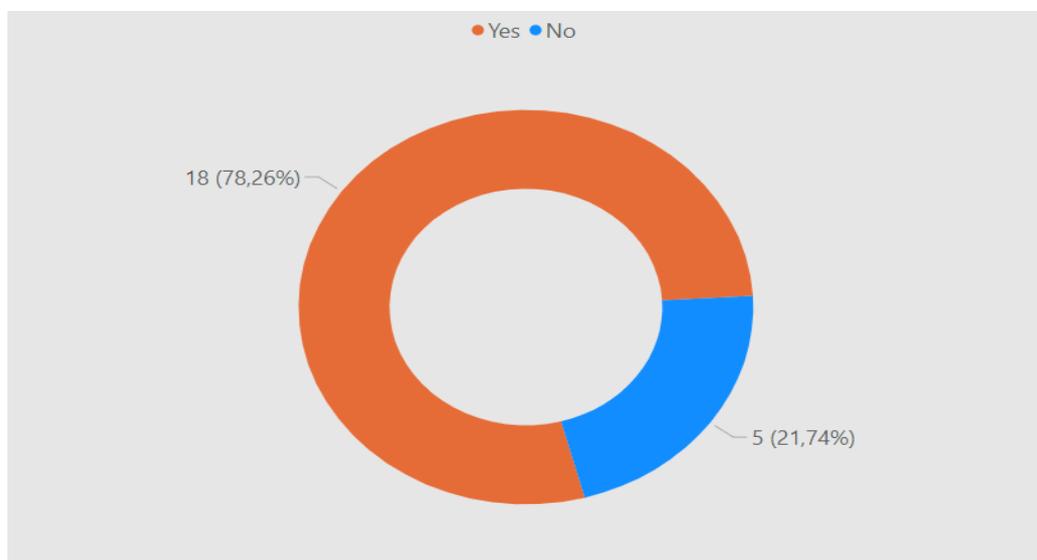

Figure 35. Perception on the recent use of GenAI tools (sample size=23).

Approximately,22% say they haven't used any type of GenAI tool compared to 78% who have used it at least once in the last 6 months.

**If not used, what would be the barriers to not adopting any GenAI tool? Figure 36.**

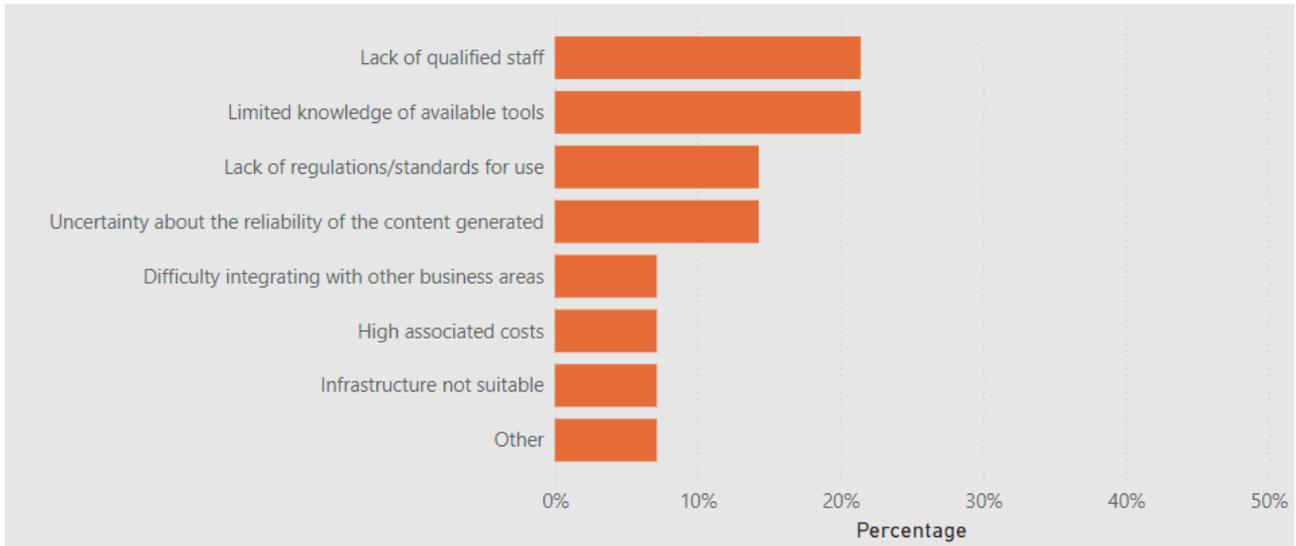

Figure 36. Barriers to not adopting GenAI tools (sample size=14).

This question has multiple answers. The top positions, tied with 21%, are associated with a *lack of qualified staff* and *knowledge of the available tools*. This was followed by 9.5% indicating a *lack of regulations/standards for use* and *uncertainty about the reliability of the content generated*.

**If you answered that you use it, which GenAI tool the company has used. Figure 37.**

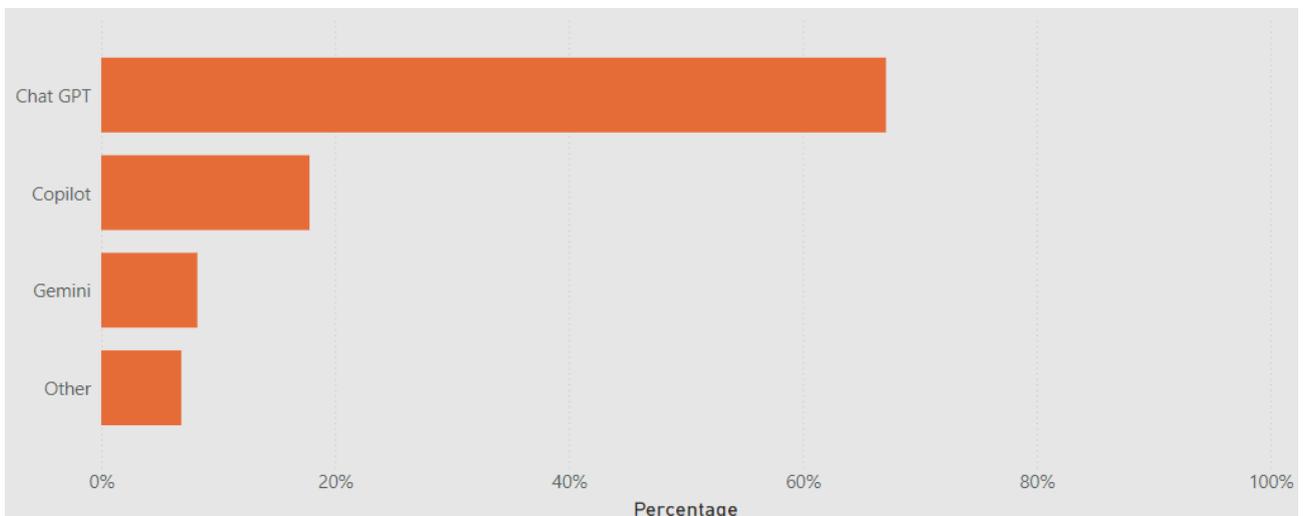

Figure 37. GenAI tools used (sample size = 17).

This question has multiple answers. The sum of the answers by all the sectors defines *Chat GPT* in first place, followed by *Copilot* and *Gemini*. *Canva* and *freepik* graphic design platforms, *TL* and *Perplexity* were also mentioned.

**Has the company or a specific employee invested in the creation of its own Generative AI tool? Figure 38.**

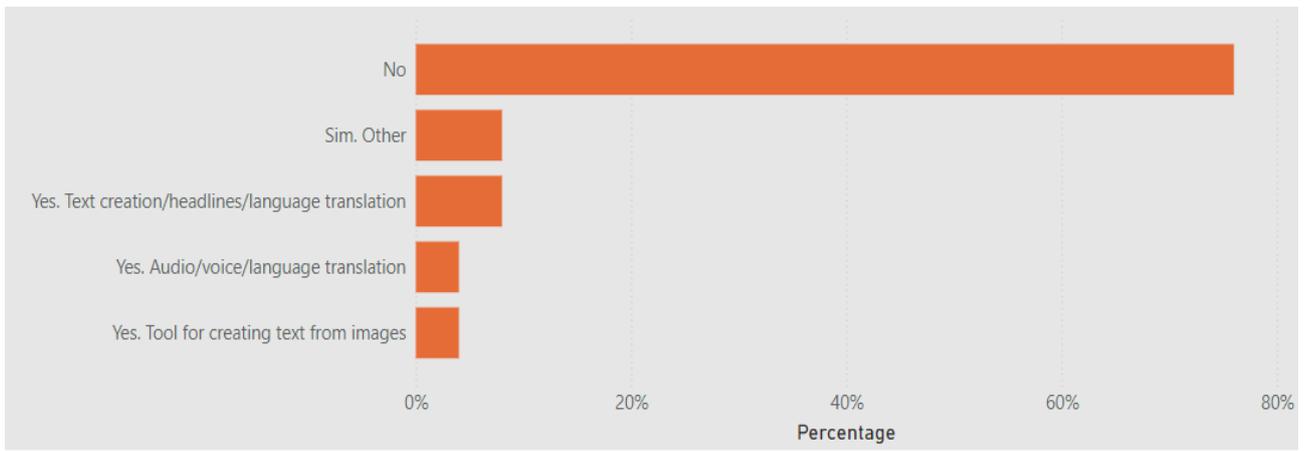

Figure 38. Creation of own GenAI tools (sample size=25)

This question has multiple answers. 76% of respondents have not invested in creating their own generative AI tool, nor have any of their employees. Tied with 8% are text creation tools, headlines, summaries, language translation tools, along with humanized customer service and marketing process automation tools listed by respondents. The group of tools for recognizing audio, voice, music and language translation were also mentioned, with a lower percentage of 4% along with tools for creating text from images

**Has the company seen any benefits from using GenAI tool(s)? Figure 39.**

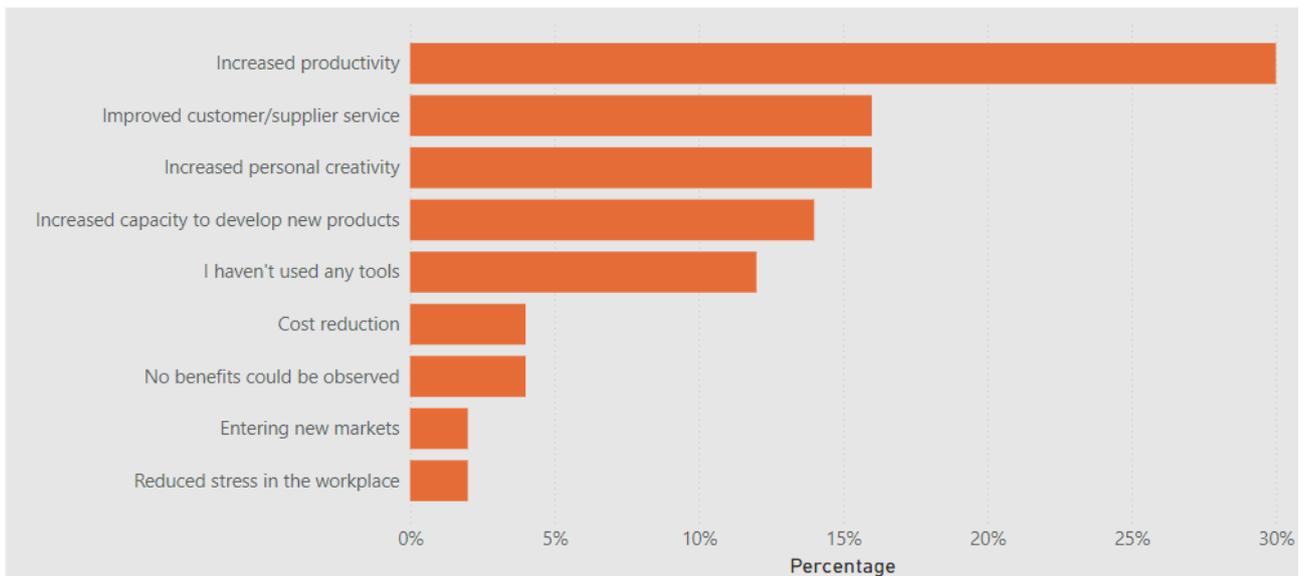

Figure 39. Benefits from GenAI tools (sample size = 50).

This question has multiple answers. The main benefit felt by employers was increased productivity, 30%, followed by improved customer service and increased personal creativity.

**What are the limitations that have contributed to the intended benefits of using GenAI not being achieved? Figure 40.**

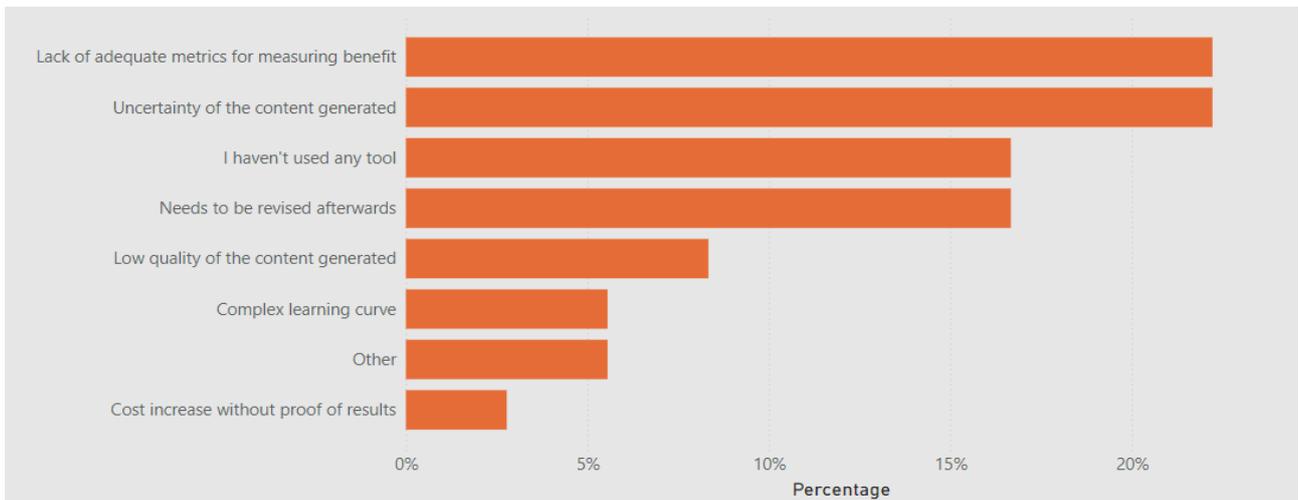

Figure 40. Limitations to achieve the goals with GenAI (sample size = 36)

This question has multiple answers. The top two with 22% each are "lack of adequate metrics to measure the benefit" and "uncertainty of the content generated". This was followed by 17% with "the need to revise it later, taking up a lot of staff time" along with the percentage of companies that didn't use any GenAI tool. The lack of rules for the security of data provided to AI tools was also mentioned.

**How does the company view the qualifications of its employees in relation to the use of Generative AI tools to help carry out work routines? Figure 41.**

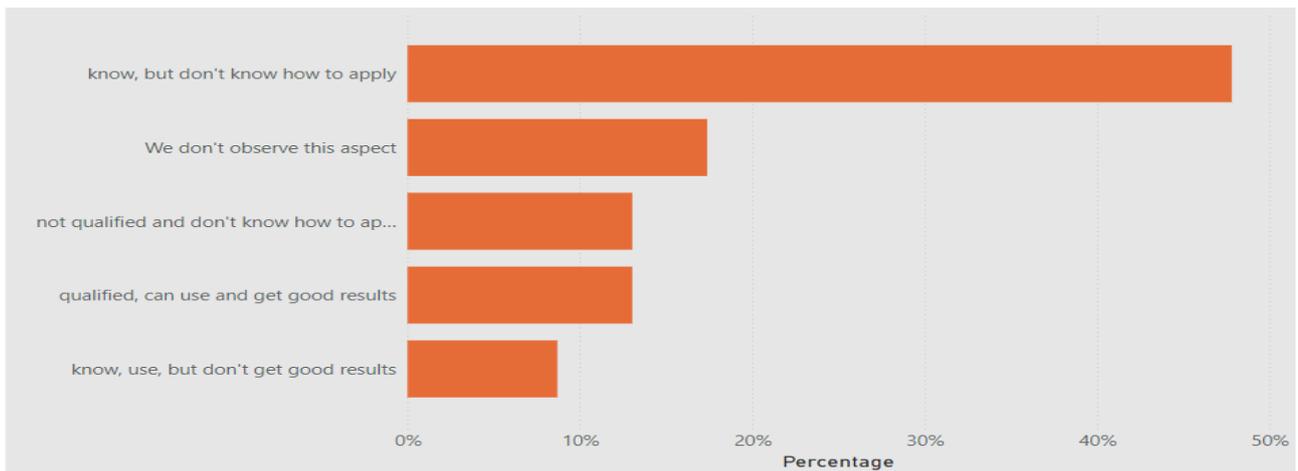

Figure 41. Qualifications of its employees (sample size 23)

For 48% of employers, most employees know about the tools, but don't know how to apply them in their daily work. On the other hand, 17% of employers do not consider this aspect relevant to their employees' skills in using GenIA in their work routines.

**Has the company already made use of a workforce composed exclusively of GenAI? Figure 42.**

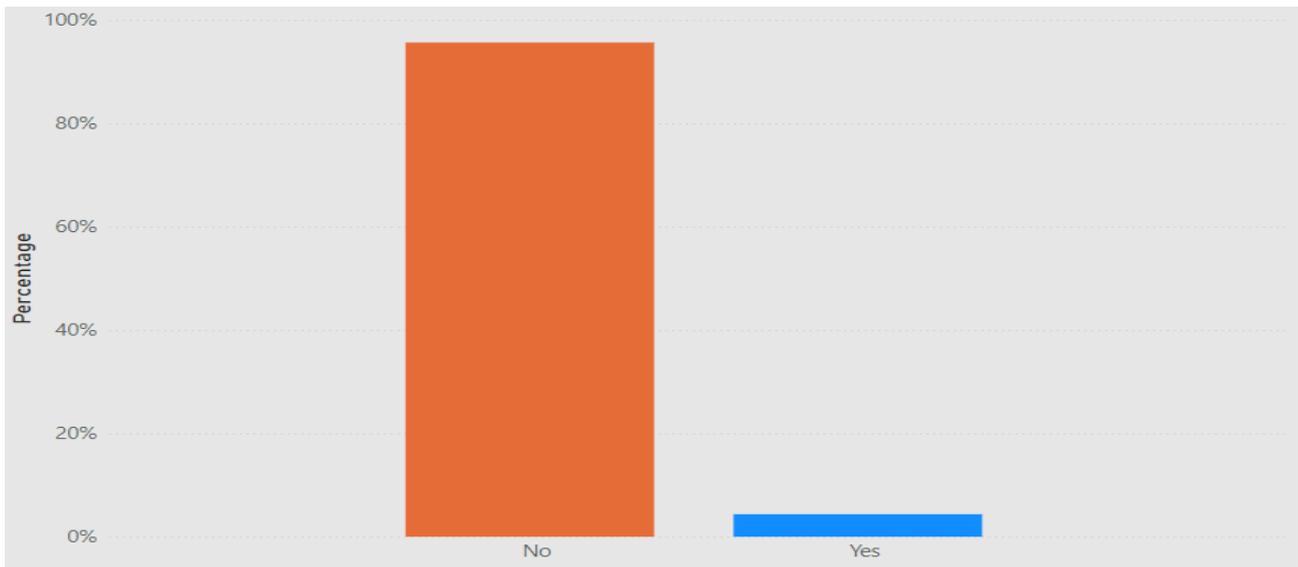
Figure 42. Workforce composed exclusively only of GenAI tools (sample size=23)

96% of companies have not used a workforce composed exclusively of GenAI. 4% indicate the existence of a full workforce based on GenAI.

**Any occupational positions (Sales, Customer Service, Graphic Design/Photography,Journalism/ContentProductionCoders/Software Developers) been impacted by the use of Generative AI? Figure 43.**

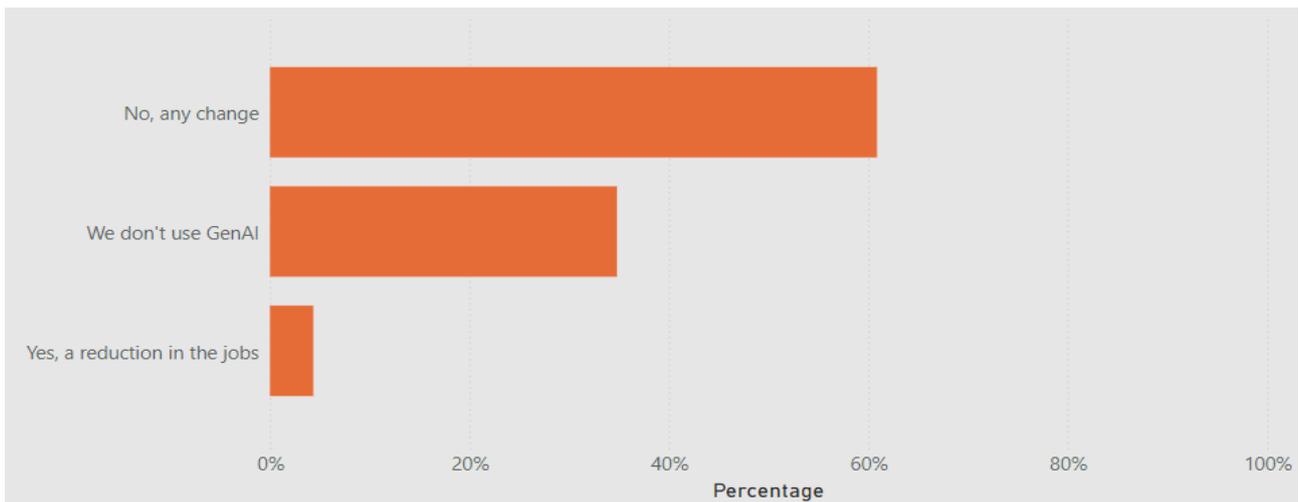
Figure 43. Positions by sector (sample size=23)

This question accepts multiple answers. 61% see no change in the number of jobs in the sectors mentioned. 4% say there has been a reduction in the number of jobs. No one reported that the use of IAGen resulted in an increase in the number of occupations for the positions surveyed

**Have any intermediate leadership positions such as managers, supervisors, coordinators been impacted by the use of Generative AI? Figure 44.**

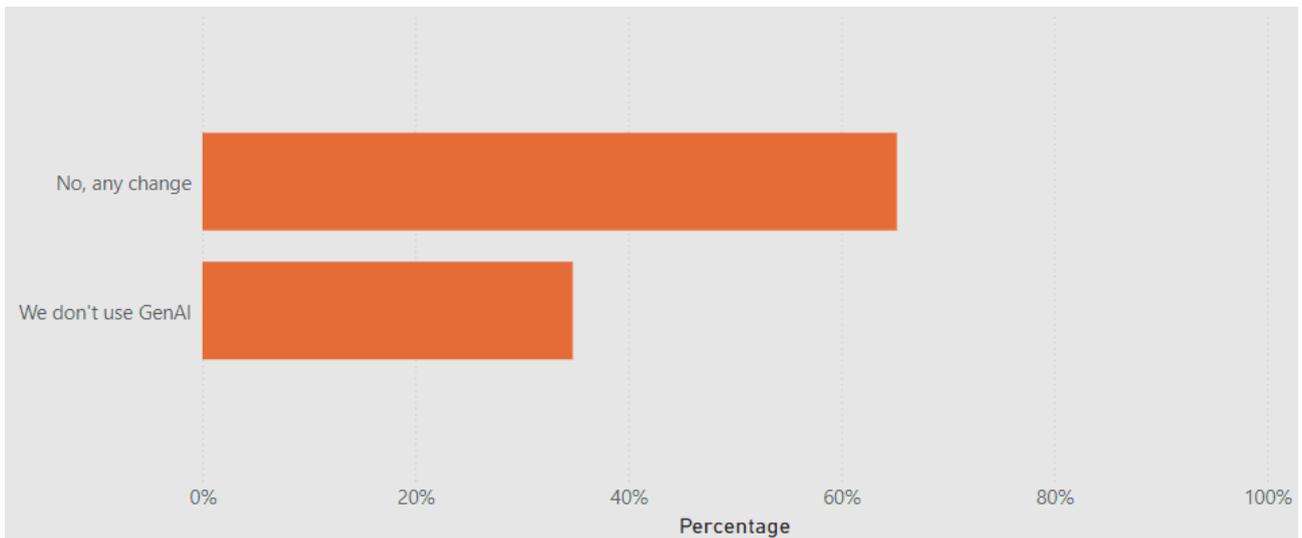

Figure 44. Intermediate leadership positions (sample size=23)

This question accepts multiple answers. In the case of intermediate positions, such as managers, supervisors and coordinators, 65% saw no change in the number of positions.

**Has the company noticed a change in the way employees carry out their activities due to the use of generative AI? Figure 45.**

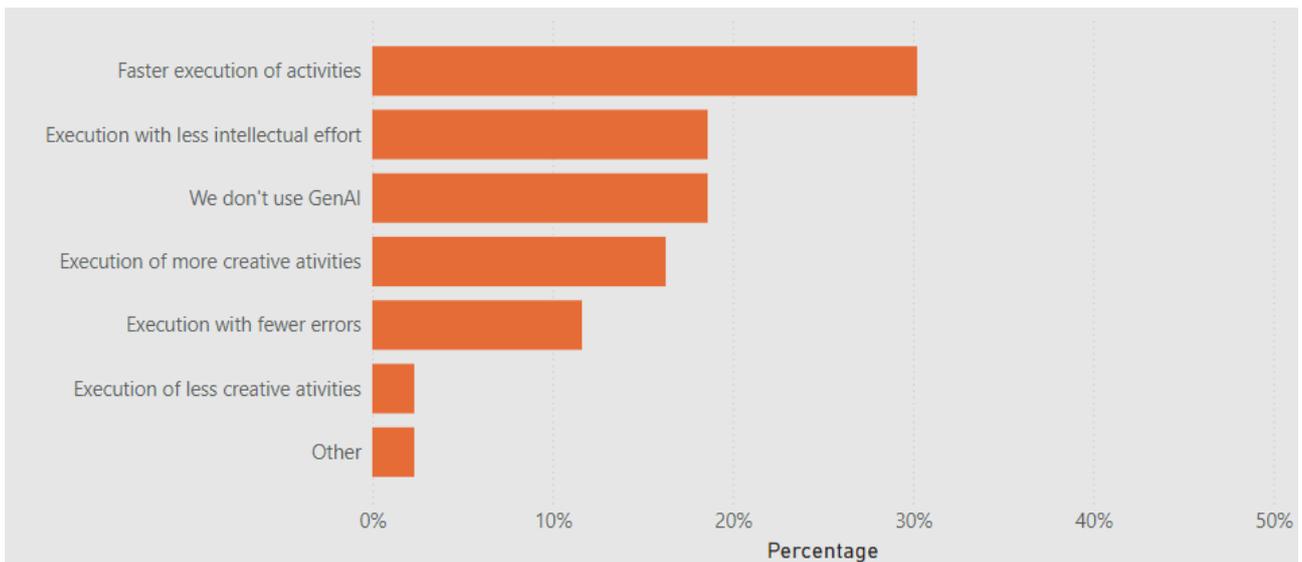

Figure 45. Change to carry out daily activities (sample size=43)

This question has multiple answers. The first position of employers' perception of patterns of change in the way their employees carry out their work was in relation to speed. 30% note that tasks are being carried out more quickly.

**Has there been any impact on payroll as a result of using Generative AI? Figure 46.**

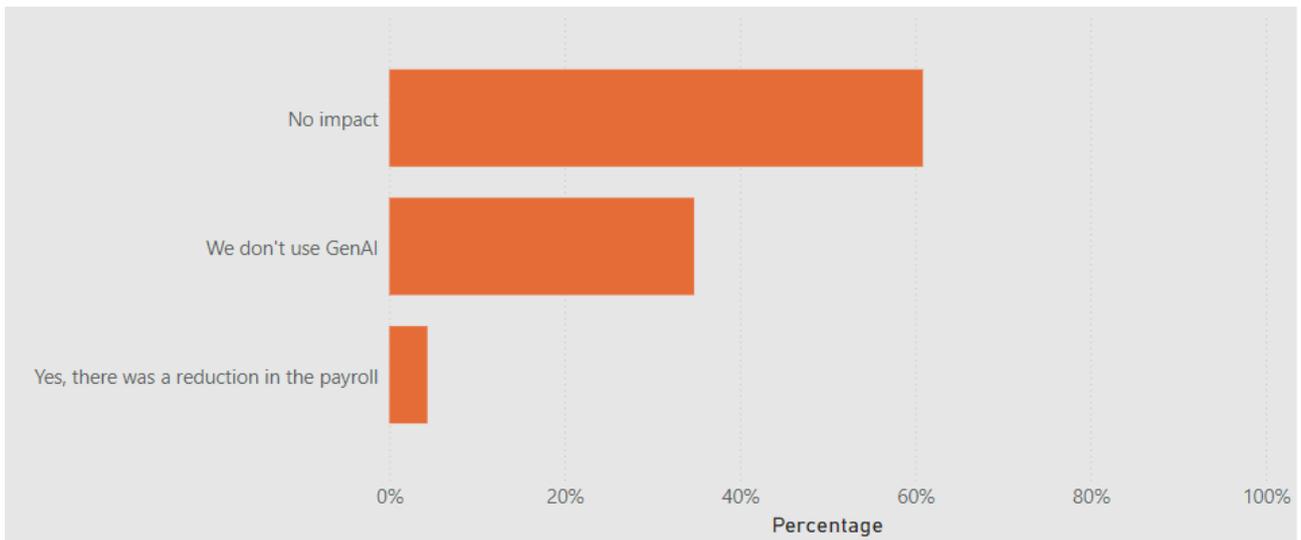

Figure 46. Impact on payroll (sample size=23)

The majority of results, 61%, felt no impact on employee payroll. 4% saw a reduction in payroll due to the use of GenAI tools.

**Does the company intend to use or expand the use of Generative AI in the next half of 2024, encompassing hardware, software, cloud computing, cybersecurity, staff training, etc? How much of the budget is earmarked for this? Figure 47.**

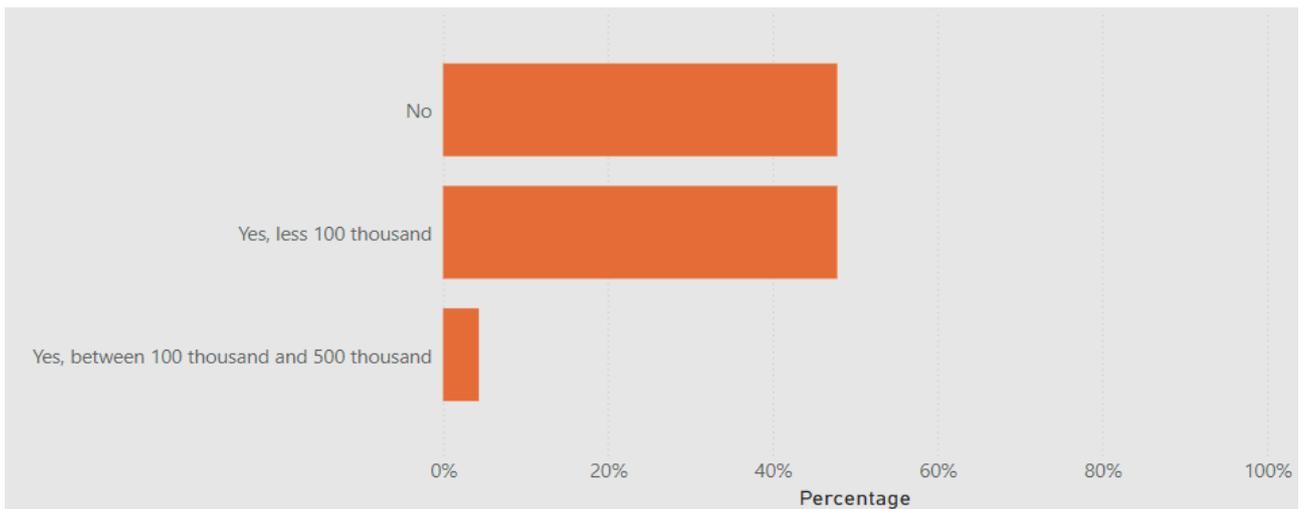

Figure 47. Budget to GenAI (sample size = 23)

The data indicates a parity in the pretension to invest or not in GenAI: 48% intend to invest amounts of less than 100,000 reais, while the same proportion say they have no budget for this.

## 4. Summary by Dimensions

### 4.1. Working hours

Among the respondents who used GenAI, Figure 28 shows that the reduction in time to carry out daily tasks was perceived as being the main modification to their routine tasks. Analysing this information, we need to regress to some conditions asked in the beginning of the survey. The form of employment contract with the greatest contribution was *CLT*, Figure 12, and *on site working hours*, Figure 13. This combination is still very widespread in the country and is noticed in Figure 14, where the majority of employees need to keep records of when they clock in and out of work, since this condition is linked to fulfilling a contract and receiving a salary. So for this fraction of employees the total hours worked do not change with GenAI use. Changes in the contractual dynamic are already being seen with the inclusion of hybrid work and remote work, which brings greater flexibility and directs time management to the worker and no longer to the employer.

For "Employers", faster execution of tasks is also observed as the first pattern of change in the activities of employees using GenAI at Figure 42. This shows an agreement from both sides, employers and employees, regarding the impact of GenAI. So, if tasks are being completed faster and the employee has to stay at the workplace to complete the contracted hours, how has this free time been operationalized? How is this evidence actually being internalized by both parties? Has this change brought benefits to both parties? Are there longer breaks between tasks? Is the employer demanding extra tasks? Is there more time available for learning and knowledge sharing? These are questions that need to be investigated, as the impact of these now subtle changes may be greater in the future.

Countries with greater investment in science and technology are already experiencing the growth of on-demand jobs, i.e. task-oriented work contracts, specially for positions requiring specialized skills. It is necessary to investigate the extent to which the on-demand approach for job contracts is in line with the interests of the workforce in Brazil.

The most valuable observation that can be reinforced is that there are numerous questions and little evidence to answer them. There is an urgent need to expand studies on the Future of Work in Brazil.

### 4.2. Working conditions

Employees working conditions can be analyzed in various ways. This includes the total working day, flexibility of entry and exit times, benefits, workload, the social environment for sharing ideas with colleagues or superiors, forms to engage the worker with the company mission among other questions scattered throughout both surveys.

Here, we will focus on Figures 19 and 20 concerning the following conditions: "Internet network with adequate speed", "computers, memory, monitor and mouse", "access to software and applications", "work desk and chair", "room lighting", and "presence of noise/parallel conversations". Technical conditions involving network speed, physical hardware and software structures are highly sought after for the scope in question. The majority of participants gave good ratings to all the items listed.

Employers were asked if they had noticed an increase in complaints from workers in the last 6 months about ergonomic conditions, the environment or equipment/software available in the workplace. Figure 48 shows that more than 80% did not notice an increase in complaints. In this sense, the two points of view are similar.

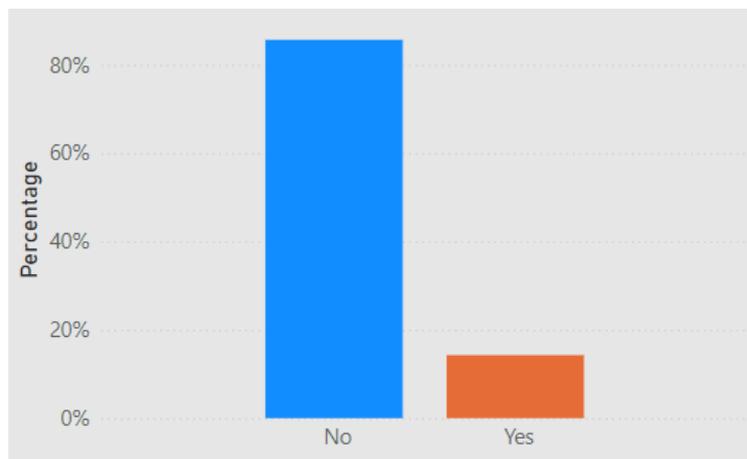

Figure 48. Perspective on increase in complaints. (Sample size = 28)

Figure 29 relates the use of GenAI to additional income or improvements in ergonomic conditions, the environment or equipment/software available in the workplace. In this sense, 11% indicated having noticed improvements in conditions, or receiving benefits or salary increases, while 59% were unable to identify any type of improvement after GenAI.

Figure 39 shows that "Employers" who used GenAI were able to identify benefits such as increased productivity, improvements in customer service, and increased personal creativity.

In summary, GenAI brings some benefits that are felt and listed by employees and employers as noticeable. These improvements have not been reflected in improvements in the daily work of employees, at least regarding the issues raised.

### 4.3. Perception of changes/inequalities

For employers, GenAI is still taking shape. This may be associated with the size of companies, as it affects the amount of available labor, investment capital, time for innovation, among other issues. Regarding the reduction in jobs, there are signs of 11% in Figure 43. This condition is also perceived by the workforce. Figures 23, 9% say there is a reduction in intermediate positions. Figure 33, positions such as manager and content producer were cited having suffered reductions attributed by GenAI.

The majority of "Employees" do not perceive any change in the number of positions for the last 6 months. However, remarks in the open questions field bring expressions like "machine in, human out". Such remarks express employees feelings about the technology. A fair assumption is that the wording employed contains fear regarding employees future with the arrival of the technology.

Open text questions provide more information from respondents.. Employer data shows that "there was a restructuring at the company with the reduction of more than 20 advisor positions". The number mentioned is considerably high for a single sector. It is therefore clear that this is a large company. This highlights the difference in reality of each business category and, consequently, of the workforce that works in it. GenAI is increasingly improving its interaction with humans to better understand user needs and provide the best answers, so positions whose tasks include providing advice (legal, financial, investment) may feel this impact. Medium or large companies with larger budgets to invest on GenAI tools will be in a better position than micro or small companies.

Even in a group with a high level of formal education, the inequality can be observed. In the question evaluating the *knowledge about the tools*: the groups that somehow have greater access to technology that have the motivation to apply the technology and that show curiosity about it, manage to integrate the content more quickly into their activities and extract the benefits, Figure 28. They report experiencing a reduction in the time it takes to carry out tasks, less intellectual efforts to perform them and increased creativity.

### 4.4. Education and skills

100% of the sampled group, Figure 10, has followed education beyond high school. Additionally, 95% have at least initiated a college degree or attained even higher degrees. Some of the job sectors surveyed require specific qualifications and skills gained through formal education. Besides, the region most represented was Southeast, the highest Human development Index (HDI) in the country.

95% of "Employers" had seen no changes in their hiring processes in the last 6 months. 5% apply changes following the indication of company members or colleagues. Those who apply changes in the hiring proccess request applicants' proof of skills through tests, and attested previous professional experience in the area of activity, reinforcing the necessity to improve skills of the workforce.

The self-perception of their own skills in Figure 22 indicates that 67% have the necessary skills to use GenAI. On the other hand, for employers, Figure 40, almost 50% identify that most workers know GenAI, but don't know how to use it. This lack of understanding between the parties hampers the fluidity of work processes.

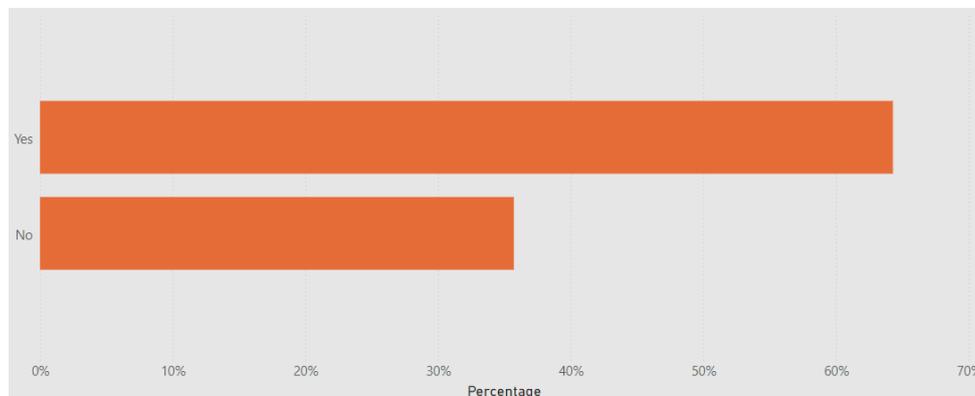

Figure 49. Training offered by the company to its employees (sample size = 23).

Figure 49, shows 65% of the companies invest in courses and lectures to develop job skills. The most recent training courses include: motivational, sales tactics, customer service, time management, quality of life and sustainability, among others. In this sense, the two responses from "Employer" and "Employee" complement each other. The skills focused on by the employer emphasize social and technical development and seek to combine sustainability. Of the 5% who say they encourage training, the focus is on GenAI tools. This corroborates the responses of the 85% of employees shown in Fig. 31.heir employers do not invest in training or lectures aimed at increasing theoretical or practical knowledge about the use of Artificial Intelligence.

The question depicted in Figure 41,raises the perception of employers about the readiness of their employees in adopting GenAI tools; the most selected answers were "limited knowledge" and "lack of qualified staff". This can also represent a view that the learning curve is too long for both employers and employees, which hinders the first steps towards learning and engaging in work activities.

As a consequence, Figure 32 shows that the majority of workers seek learning on their own, mainly using the internet. This attitude is welcomed and required, but the task of broadening knowledge cannot be directed solely at workers, as this would increase the inequality that already exists. Without guidance, it causes misalignment and can be time-consuming to find something valuable. Knowledge needs to be universalized and accessible. Creating a strong society and being able to deal with the changes generated by the digital transformation, requires the contribution of the entire production chain, including the transversal sectors.

### 4.5. Wage

Income is a delicate factor for Brazilians, considered to be a very personal subject. Thus, to measure this indicator, the first approach was established by bands, Figure 11. The remaining questions were indirectly assessed by composing information on benefits, Figure 17, and extra income, Figure 18. The latter is associated with motivation and we can see that the majority of respondents report that they don't receive extra income and would feel more motivated if they were offered performed-based bonuses. These questions were investigated again in the third module of the questionnaire to identify the impact of IA on this parameter. Figure 29 associates additional income with improved conditions in the workplace. Of the participants who used GenIA, the majority do not receive additional income.

Figure 46 shows whether there was an impact on the payroll of the "Employers".The majority of responses was that employers felt no impact at all, with a minority saying that they had seen a reduction in payroll with the use of GenAI. Figure 47 shows the future intentions of preparing a budget to invest or not in this technological innovation. We can observe two very distinct groups, those who do not set aside a budget for this endeavor and those who want to accept the risk of innovation. The majority of investments in the range of up to one hundred thousand reais (R$100k) is in line with the profile of the companies involved. There is a minority that will set aside a larger part of the budget (R$ 100K to 500K).

The employer's decision has an impact on the employee's future, so it is important that the needs of both are transparent. GenAI, centered on the human being, participates in the work environment as an instrument forming a triad to achieve the desired objectives.

The general assessment of the future for employees and employers in the investigated sectors is that GenAI tools will help workers. They recognize the need to adapt to change and learn to deal with technology. They seek information and need guidance to better achieve their goals. At the same time, they are insecure and looking for direction.

What the analyses highlight is the need to combine the efforts of all those involved in the production chain, governmental and non-governmental organizations, associations and affiliations with political influence, employers and employees in order to go through this transformation while managing the impacts.

### 5. Opportunities and Lessons Learned

We are presenting the initial observations of this research and, as we delved deeper into the data, we identified that more elaborate explorations could provide more information about the different points of view of employees and employers. In addition, we evaluated the issue of disaggregating the data by category and its relevance in relation to the overall assessment.

The methodological approach adopted allows the study to be replicated, extended or focused on other niches of interest. It can cover a wider range of sectors and can certainly be improved for the next assessment.

We understand that stakeholders need to be heard and we would like to expand the scope of the research to encompass all regions of the country. We aim not only to reduce statistical uncertainty but also to truly identify the different needs geographically located. To do this, we will need support that goes beyond our initial partnerships with associations, confederations and unions.

Decision-making to address the challenges associated with the expansion of IAGen can be better informed when supported by real results from research. This type of approach aligns objectives and tends to improve the effectiveness of the actions undertaken.

## 6. Conclusions

GenAI tools are increasingly permeating the activities in the society. The transformations motivated by it are taking place in a subtle way, especially in micro and small businesses. This sample survey reveals the perception of employers and employees on five sector of the workforce in Brazil. The results are analyzed in light of six analytical workforce dimensions.

The exploratory methodology involved administering a questionnaire to a sample of the target population, with distinct versions tailored for employers and employees. Ensuring data security and respondent anonymity fostered participant trust and confidence, thereby encouraging candid and uninhibited responses. From a technical standpoint, the standardization of responses facilitated uniform analysis, while the inclusion of open-ended questions provided valuable insights into aspects not explicitly addressed by the closed-ended items.

Based on the analytical dimensions employed in this survey, we can highlight several key takeaways from this work:
- Employees have identified a reduction in time spent on work tasks as a noticeable effect of adopting generative AI (GenAI) in their activities. However, this benefit may not be as appealing to those working under fixed working hours regimes.
- Employees in the professional occupations surveyed report experiencing satisfactory working conditions. This environment could facilitate the integration of generative AI (GenAI) into their routines; however, such adoption has not yet occurred;
- Approximately 11% of respondents reported indications of workforce reductions. Analysis of open-ended responses revealed a sense of apprehension regarding the introduction of generative AI (GenAI), with employees expressing concerns about potential job displacement and the broader implications of AI integration into their work environments;
- The need for employee training in new technologies is not yet a priority for the majority of companies participating in the survey. Consequently, employees utilizing these technical tools often resort to self-directed learning, frequently outside of their regular working hours;
- This self-driven adoption often occurs in the absence of structured training or clear organizational policies.The lack of employer-provided training and incentives may stem from various factors, including uncertainty about the technology's benefits, concerns over data security, or the absence of a clear implementation strategy. Nonetheless, this gap presents an opportunity for organizations to support their workforce by providing access to GenAI tools, offering training programs, and establishing clear usage policies;
- A few companies have suggested interest in adopting GenAI as part of the work tasks in the newer future. This is a positive sign about what may come in the next few years in small and medium companies in Brazil.

Considering that this survey was modeled and developed in a short period of time the structure fully met the objective. Based on the attained experiences some specific refinements can be done. It will also be necessary to improve the dissemination phase of the forms, expand partnerships, establish some form of reward to encourage engagement and thereby achieving stronger statistical representativeness for all sectors.

The growth and sustainability of society is the result of the labor market. The future of the social organization called "work" and of the worker is in transition. We hope that the evidence observed in this research can contribute to the creation of a joint plan involving leadership organizations and political force. Actions that prioritize human rights, ethics, justice and, at the same time, maximize the development of work processes need to be disseminated equally so that the workforce can cope with the changes.

The IIA-LNCC supports open research and is available to share the results of this study. We hope to contribute to the development of the future of work, where personal skills are combined with technological innovation, paving the way for enriching paths for all sectors involved.

# Methodological Annex

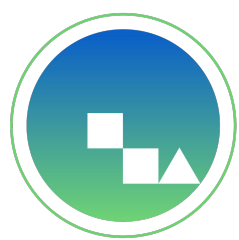

# APPENDIX I

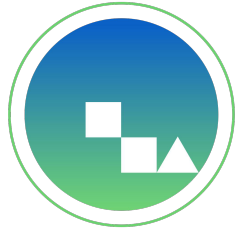

# SURVEY: THE IMPACT OF USING GENERATIVE AI AT WORK FOR EMPLOYERS IN THE FOLLOWING SECTORS:

Sales | Customerservice | Graphic Design/Photography

Journalism/Content Production | Coders/Software Developers

## A) Identification     (mandatory answers)

**1. CPF:**

**2. Age:**

**3. Education Level:**
(  ) Incomplete primaryeducation
(  ) Complete primaryeducation
(  ) Incomplete  high school
(  ) Completed high school
(  ) Incomplete highereducation
(  ) Complete highereducation
(  ) Postgraduate or Extension

**4. City:**

**5. Company name:**

**6. Position in the company:**

## B) General Scope

7. **What form of employment relationship is primarily used for each of the sectors? Select the form of employment most applicable to your company for each of the sectors. Or use the "Other" field for your own description.** (Single answer for (1,2,3,4,5) and mandatory).

   | Sales | Customer service | Design/ photography | Journal/content | Software development |
   |---|---|---|---|---|
   | ( )CLT | ( )CLT | ( )CLT | ( )CLT | ( )CLT |
   | ( )PJ | ( )PJ | ( )PJ | ( )PJ | ( )PJ |
   | ( )Temporary | ( )Temporary | ( )Temporary | ( )Temporary | ( )Temporary |
   | ( )Trainee | ( )Trainee | ( )Trainee | ( )Trainee | ( )Trainee |
   | ( )Other | ( )Other | ( )Other | ( )Other | ( )Other |

8. **What forms of work has the company adopted in the last 6 months?**
   **(Singleanswer for (1,2,3) and mandatory).**

   | Sales | Customer service | Design/ photography | Journal/content | Software development |
   |---|---|---|---|---|
   | ( )On site | ( )On site | ( )On site | ( )On site | ( )On site |
   | ( )Teleworking | ( )Teleworking | ( )Teleworking | ( )Teleworking | ( )Teleworking |
   | ( )Hybrid | ( )Hybrid | ( )Hybrid r | ( )Hybrid | ( )Hybrid |

9. **Does the company use any tools such as: time cards, time sheets, electronic controls or others to monitor workers' attendance and time?**
   ( ) Yes
   ( ) No

10. **What tool(s) does the company use to retain talent? If there is no corresponding answer, use the "Other" field for your own description.** (Multiple answer and mandatory).
    ( ) Individual meetings with managers (reviewof goals and challenges)
    ( ) Individual meetings with HR manager(evaluation of plans and competencies)
    ( ) Mentoring by more experienced professionals with junior professionals
    ( ) Encouragement to send e-mails or messages about ideas/improvements to managers and/or HR
    ( ) Anonymized suggestion box
    ( ) Other: _______________________________

11. **Does the company invest in refresher courses/lectures for its employees? If the answer is "Yes", indicate in the "Other" field what was the last training/lecture (in person or online/internal or external) that the company made available to its employees.** (Multiple answerand mandatory).
   ( ) Yes
   ( ) No
   ( ) Other: _______________________

12. **Does the company usually provide extra remuneration or awards linked to an increase overall productivity or productivity by sector?** (Single, compulsory answer).
   ( ) Yes
   ( ) No
   ( ) Eventually

13. **Has there been any kind of change in the hiring format in the last 6 months?** (Singlemandatory answer)
   ( ) Yes
   ( ) No

14. **If answer 13 was "Yes", indicate the factor that led to the change in the contracting format. There can be more than one answer. Use the "Other" field for your own description** (Multiple answer and not compulsory).
   ( ) proof of academic trainingrequired
   ( ) proven previousprofessional experience in the fieldof work
   ( ) proof of specificskills in the area in question by means of applied tests
   ( ) specific certifications in the area in question
   ( ) referral from company members
   ( ) Other: _______________________

15. **In the last 6 months, has the company noticed an increase in requests from workers about difficulties accessing the internet, problems with hardware devices (computers, mouse, headphones/microphones), problems with software (applications in use), problems with the environment (desk, chair,lighting, noise) that reflect on barriers to their working day?** (Single mandatory answer).
   ( ) Yes
   ( ) No

## C) Specific Scope

16. **Which sentence below best describes the company's vision of Generative AI tools. Use the "Other" field for your own description (Single answer required).**
    ( ) We have no knowledge of Generative AI tools
    ( ) We hear about Generative AI, but we have no contact with it in the company
    ( ) We have no interest in using Generative AI in any sector of the company
    ( ) I know that other companiesin the same sector alreadyuse it, but I don't know how to implement it
    ( ) We'd like to use it, butI don't know how to start
    ( ) I alreadyuse Generative AI
    ( ) Other: ______________________

17. **In the last 6 months, have any of these sectors (Sales, Customer Service, Graphic Design/Photography, Journalism/Content Production, Software Coders/Developers) used a Generative Artificial Intelligence tool, even in the testing phase?**
    ( ) Yes
    ( ) No

18. **If answer 17 was "No", what would be the barriers to not adopting any tool? There may be more than one answer. Use the "Other" field for your own description (Multiple answer and not mandatory).**
    ( ) Limited knowledge of available tools
    ( ) Uncertainty about the reliability of the content generated
    ( ) Lack of regulations/standards for use
    ( ) Lack of qualified staff
    ( ) Infrastructure not suitable
    ( ) Difficulty integrating with other businessareas
    ( ) High associated costs
    ( ) Other: ______________________

19. **If the answer was "Yes", indicate which sector and which Generative AI tool the company has used. There may be more than one answer. Use the "Other" field for your own description.** (Multiple answer and not mandatory).

| Sales | Customer service | Design/ photography | Journal/content | Software development |
|---|---|---|---|---|
| (  )Chat GPT | (  )Chat GPT | (  )Chat GPT | (  )Chat GPT | (  )Chat GPT |
| (  )Gemini | (  )Gemini | (  )Gemini | (  )Gemini | (  )Gemini |
| (  )Copilot | (  )Copilot | (  )Copilot | (  )Copilot | (  )Copilot |
| (  )Looka | (  )Looka | (  )Looka | (  )Looka | (  )Looka |
| (  )Night Café | (  )Night Café | (  )Night Café | (  )Night Café | (  )Night Café |
| (  ) Other | (  ) Other | (  ) Other | (  ) Other | (  ) Other |

20. **Has the company or a specific employee invested in the creation of its own Generative AI tool? If the answer is "Yes", indicate in the "Other" field the area(s) of activity the tool covers. There may be more than one answer.** (Multiple answers required).

   (  ) No

   (  ) Yes. Text creation/headlines/summaries/language translation tool

   (  ) Yes. Audio/voice/music recognition/language translation tool

   (  ) Yes. Photo image recognition tool

   (  ) Yes. Tool for creating text from images

   (  ) Yes. Tool for creating images from text

   (  ) Yes.  Other: ______________________

21. **Has the company seen any benefits from using its own Generative AI tool(s) or those available on the market? There may be more than one answer. Use the "Other" field for your own description.** (Multiple answers required).

   (  ) Increased productivity

   (  ) Cost reduction

   (  ) Reduced friction/stress in the workplace

   (  ) Increased personalcreativity

   (  ) Reduction of environmental impact

   (  ) Improved customer/supplier service

   (  ) Increased capacity to develop new products

   (  ) Entering new markets

   (  )  Other: ______________________

   (  ) No benefitscould be observed

   (  ) I haven'tused any Generative Artificial Intelligence tools

**22. Indicate the limitations that have contributed to the intended benefits of using Generative Artificial Intelligence not being achieved. There may be more than one answer. Use the "Other" field for your own description.** (Multiple answers required).

( ) Low qualityof the content generated

( ) Uncertainty of the content generated

( ) Complex learning curve, requiring a lot of additional time and effort from the team

( ) Needs to be revisedafterwards, taking up a lot of staff time

( ) Cost increase withoutproof of results

( ) Lack of adequate metrics for measuring benefit

( ) Other: _______________________________

( ) I haven'tused any Generative Artificial Intelligence tools

**23. How does the company view the qualifications of its employees in relation to the use of Generative AI tools to help carry out work routines?** (Single, compulsory answer).

( ) Most employees are already qualified, can use the tools and get good results.

( ) Most employees are not qualified and don't know how to applythe tools.

( ) Most employees know about the tools but don't know how to apply them in their daily work.

( ) Most employees know and use the tools, but they don't get good results.

( ) Most employees show no interest in dealing with these tools

( ) We don't observe this aspect

**24. Has the company already made use of a workforce made up exclusively of Generative AI?** (Single, compulsory answer).

( ) Yes

( ) No

25. **Any occupational positions (Sales, Customer Service, Graphic Design/Photography, Journalism/Content Production, Coders/Software Developers) been impacted by the use of Generative AI? There may be more than one answer. If the answer is "Yes", please indicate in the "Other" field which position(s) you occupy.** (Multiple answer and mandatory).
    ( ) Yes, there was an increase in the number of jobs
    ( ) Yes, there was a reduction in the number of jobs
    ( ) No, any change
    ( ) We don't use Generative AI
    ( ) Other: ______________________

26. **Have any intermediate leadership positions such as managers, supervisors, coordinators been impacted by the use of Generative AI? There may be more than one answer. If the answer is "Yes", indicate which position(s) in the "Other" field.** (Multiple answer required).
    ( ) Yes, there was an increase in the number of positions
    ( ) Yes, there was a reduction in the number of positions
    ( ) No, any change
    ( ) We don't use Generative AI
    ( ) Other: ______________________

27. **Has the company noticed a change in the way employees carry out their activities due to the use of generative AI? Check the corresponding boxes or use the "Other" box for your own description. There may be more than one answer.** (Multiple answers required).
    ( ) Faster execution of activities
    ( ) Slower execution of activities
    ( ) Execution of activities with greater intellectual effort
    ( ) Execution of activities with less intellectual effort
    ( ) Execution of activities with more errors
    ( ) Execution of activities with fewer errors
    ( ) Execution of more creative activities
    ( ) Execution of less creative activities
    ( ) No change in pattern can be observed
    ( ) We don't use Generative AI
    ( ) Other: ______________________

**28. Has there been any impact on payroll as a result of using Generative AI? (Single, compulsory answer).**

( ) Yes, there was an increase in the payroll

( ) Yes, there was a reduction in the payroll

( ) No impact

( ) We don't use Generative AI

**29. Does the company intend to use or expand the use of Generative AI in the next half of 2024, encompassing hardware, software, cloud computing, cybersecurity, staff training, etc? How much of the budget is earmarked for this? (Single mandatory answer).**

( ) Yes, more than 500 thousand

( ) Yes, between 100 thousand and 500 thousand

( ) Yes, less 100 thousand

( ) No

**30. Use the space below to describe any need or difficulty related to accessing or using Generative AI tools that the company identifies. (Textual answer and not mandatory).**

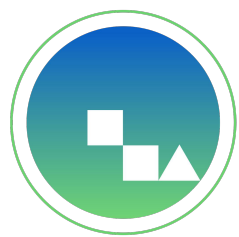

# APPENDIX II

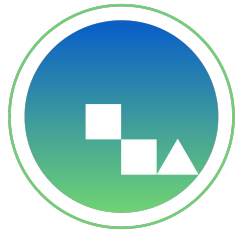

## 2. Technical Methodology

Alguns aspectos mais técnicos sobre a construção da metodologia, são apresentados nesta seção.

### 2.1 Steps for the survey construction methodology

Question form – In this sense, the definition of a direct question is attributed, as a form of response being "Yes" or "No", a multiple choice question where there could be more than one possible answer, a simple choice question or an open question with a field of free description.

Type of questions - refers to whether or not their selection on the form is mandatory.

Response form – Direct response choice form presented with described options, responses with open field for free description

Response Type – Acceptable single response field, multiple responses or open descriptive.

Number of questions – Determine the limit between the number of questions sufficient to achieve the research objective and the minimum time required for the participant to complete them.

Ordering the questions – Identify the best structure so that filling out the questions is fluid and not tiring.

Flow of ideas, cohesion and engagement - Since participation is voluntary and there is the possibility of withdrawing at any time, the respondent needs to be engaged with the topic, feel involved with the questions in order to remain active until the end of the three sections.

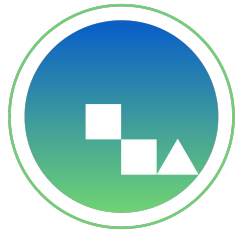

Consistency with the study objectives – The questions must capture the participant's reality in relation to the observed theme.

Time and agility in accessibility – Forms must be available in an accessible way across multiple devices, operating systems and internet browsers. They must be scalable for multiple connections in order to remain active 24 hours/day throughout the entire research period.

Relevance to the respondent – The survey should spark the participant's interest so that they feel their contribution is valuable.

## 2.2 Premises for constructing questions and answer alternatives
- Ease of understanding - questions and answers containing words used in common sense in everyday language.
- Directivity - minimize the possibility of interpretability in reading questions and answers.
- Exhaustive description of rules in the questions – repetition of instructions in the header to clarify the need and streamline completion.
- Most common alternatives for single or multiple answer choices – helps reduce completion time and the respondent's line of reasoning.
- Encouragement of own responses through open fields for completion – the respondent is not limited to the available alternatives and can express their opinion freely.
- Estimated response time between 8 and 10 min.

## 2.3 Database Modeling
The IIA developed the application using the Laravel 8 framework, MySQL database and the Bootstrap 5 front-end framework. Laravel was used for its efficiency in back-end applications, while Bootstrap 5 was chosen to ensure an attractive visual design and a user-friendly, responsive interface that is compatible with different devices.

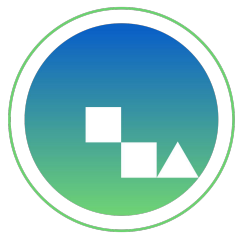

The server of the LNCC's Data Extreme Lab (DEXL-Lab) dedicated to data science received a dedicated space for this application, in addition to extra layers of security ensuring, in addition to the necessary privacy, the scalability of simultaneous accesses.

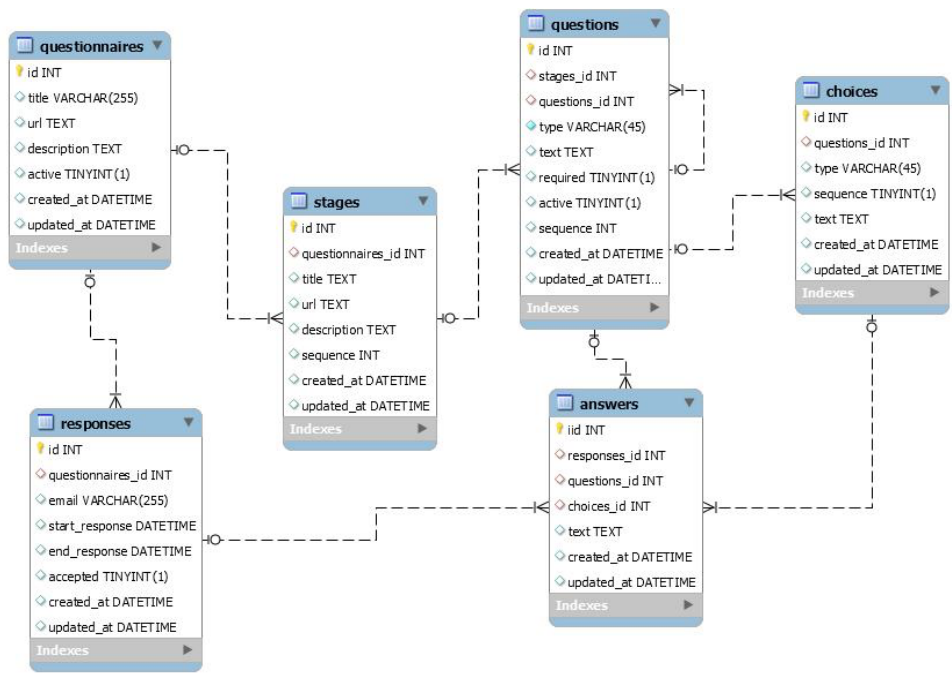

The form was designed with a modular structure, divided into three sections. At the end of each section, the respondent was immediately directed to the next section, making it more dynamic to complete. Several validation mechanisms were implemented, one of which was to ensure the uniqueness of the respondent through a personal coder, aiming to obtain answers that reflected the participant's reality and minimising false entries. The form could only be completed once by including an existing personal validator. Another mechanism worth mentioning concerns participant engagement and completeness of answers. If any mandatory question was not completed due to distraction, the screen displayed a warning message and automatically redirected to the incomplete section. Progress was only allowed after checking for completeness. Participants could also leave the form at any time by leaving the page.